\documentclass{nature}
\usepackage[pdftex]{graphicx}
\usepackage[pdftex,bookmarks,colorlinks,breaklinks]{hyperref} 
\usepackage{amsmath,amssymb}
\usepackage{mathtools}
\hypersetup{linkcolor=red,citecolor=blue,filecolor=dullmagenta,urlcolor=blue} 
\hypersetup{linkcolor=red,citecolor=blue,filecolor=blue,urlcolor=blue} 
\renewcommand{\eqref}[1]{(\ref{#1})}
\usepackage{lineno}
\usepackage{caption}

\newcommand{\tp}[1]{\textcolor{red}{#1}}

\makeatletter
\let\saved@includegraphics\includegraphics
\AtBeginDocument{\let\includegraphics\saved@includegraphics}
\renewenvironment*{figure}{\@float{figure}}{\end@float}
\makeatother

\begin{document}

\title{Enhanced Signal-to-Noise Performance of EP-based Electromechanical Accelerometers}


\author{Rodion Kononchuk$^1$, Jizhe Cai$^2$, Fred Ellis$^1$, Ramathasan Thevamaran$^2$, Tsampikos Kottos$^1$}
\maketitle

\begin{affiliations}
\item Wave Transport in Complex Systems Lab, Department of Physics, Wesleyan University, Middletown, CT-06459, USA
\item Department of Engineering Physics, University of Wisconsin-Madison, Madison, WI-53706, USA
\end{affiliations}

\begin{abstract}
Exceptional points (EP) are non-Hermitian spectral degeneracies where both eigenvalues and their corresponding eigenvectors 
coalesce \cite{K13,B04,MA19,PLBKC21}. Recently, EPs have attracted a lot of attention as a means to enhance the responsivity 
of sensors, via the abrupt resonant detuning occurring in their proximity \cite{W14,HHWGGCK17,COZWY17,CSHCCGA18,DLYQH19,
HSCK19,LLSYV19,XLKA19,ZSHYSJ19,W20,YWWGV20,KK20,HNCHKLLK20,LC18,L18,W20a}. In many cases, however, the EP 
implementation is accompanied by noise enhancement leading to the degradation of the signal-to-noise performance of the sensor
\cite{LLSYV19,W20,YWWGV20,LC18,L18,W20a}. The excess noise can be of fundamental nature (due to the eigenbasis collapse) 
or of technical nature associated with the use of amplification mechanisms utilized for the realization of EPs. Here we show, using 
an EP-based ${\cal PT}$-symmetric \cite{Bender,LRC} electromechanical accelerometer, that the enhanced technical noise can be 
surpassed by the enhanced responsivity to applied accelerations. The noise due to eigenbasis collapse is mitigated by exploiting 
the detuning from a transmission peak degeneracy (TPD), which forms when the sensor is weakly coupled to transmission lines, as 
a sensitivity measurant. These TPDs occur at a frequency and controlled parameters for which the bi-orthogonal eigen-basis is still 
complete and are distinct from the EPs of the ${\cal PT}$-sensor. They also show a slightly enhanced detuning rate compared to the 
typically utilized EPs. Our device demonstrates a three-fold signal-to-noise ratio enhancement compared to configurations for which 
the system operates away from the TPD.    
\end{abstract}

Exceptional points (EP) are spectral degeneracies occurring in the parameter space of an open (non-
Hermitian) system where two $N=2$ (or more) eigen-frequencies and their corresponding eigenvectors coalesce \cite{
K13,B04,MA19,PLBKC21}. Recently, their importance as a means to design hypersensitive sensors has been highlighted 
by a number of 
contributions \cite{W14,HHWGGCK17,COZWY17,CSHCCGA18,LC18,DLYQH19,HSCK19,LLSYV19,XLKA19,ZSHYSJ19,
W20,YWWGV20,KK20,HNCHKLLK20}. At the core of these proposals is the possibility to induce a sublinear detuning of 
the resonant frequencies from the EP degeneracy when the system interacts with a perturbing agent. Specifically, in the 
proximity of an $N-$th order EP, the degenerate resonances abide to Puiseux generalized expansions leading to a 
resonance detuning $\Delta f\equiv\left|f-f_{\rm EP}\right|$ that follows an $N-$th root behavior $\Delta f\sim \varepsilon^{
1/N}$ with respect to the perturbation strength $\varepsilon$ imposed to the system by the presence of the perturbing 
agent. Obviously, in the small perturbation limit this sublinear response signifies an enhanced sensing as compared to 
a linear response, i.e. $\Delta f\sim \varepsilon\ll \varepsilon^{1/N}$, utilized in many sensing schemes that rely on Hermitian 
degeneracies \cite{F16,X12,B00,AH13}. In fact, the proposed EP-based protocols have additional advantages compared 
to other sensor schemes (e.g. slow light) whose operational principle relies on abrupt intensity variations of the measured 
signal. These type of sensors turned out to be extremely sensitive at the expense of their dynamic range \cite{FBG1,SKOLIANOS2}. 
The dynamic range is an important metric in the characteristics of the sensor, and it is defined as the ratio between the maximum 
and the minimum perturbation that a sensor can measure.

Given that efficient sensing is of importance in many fields, the emerging idea of boosting the sensitivity of a 
particular system via non-Hermitian degeneracies could have substantial ramifications across several technological areas. 
It is therefore paramount to analyze various aspects of EP-sensing and deduce the sensing performance of such sensors. 
While increased sensor responsivity has been demonstrated in several EP-based sensors \cite{HHWGGCK17,COZWY17,
CSHCCGA18,DLYQH19,HSCK19,HNCHKLLK20,LLSYV19}, the signal-to-noise performance has been controversially 
debated in recent theoretical studies \cite{XLKA19,LC18,ZSHYSJ19,W20,L18,W20a}. Obviously, it is imperative to confront 
these disagreements via a direct experimental investigation of the effects of noise in EP-sensor performance and identify 
platforms and conditions (if any) under which they demonstrate superior performance. Currently, the only experiment 
that has analyzed the effects of noise in the EP-sensitivity (precision of a sensor) has utilized a Brillouin ring laser gyroscope 
(RLG) platform \cite{LLSYV19,YWWGV20}. The measurements invoked ultranarrow spectral line, highly suitable for precise 
measurements of small frequency shifts. Unfortunately, the conclusions of this study were discouraging as far as the 
performance of EP-sensing is concerned: the expected boost in responsivity of the gyroscope turned out to be limited by a 
broadening of the two laser linewidths due to enhanced noise effects associated with the collapse of the eigenmode basis.

\begin{figure}
\centering
\includegraphics[width=1\columnwidth]{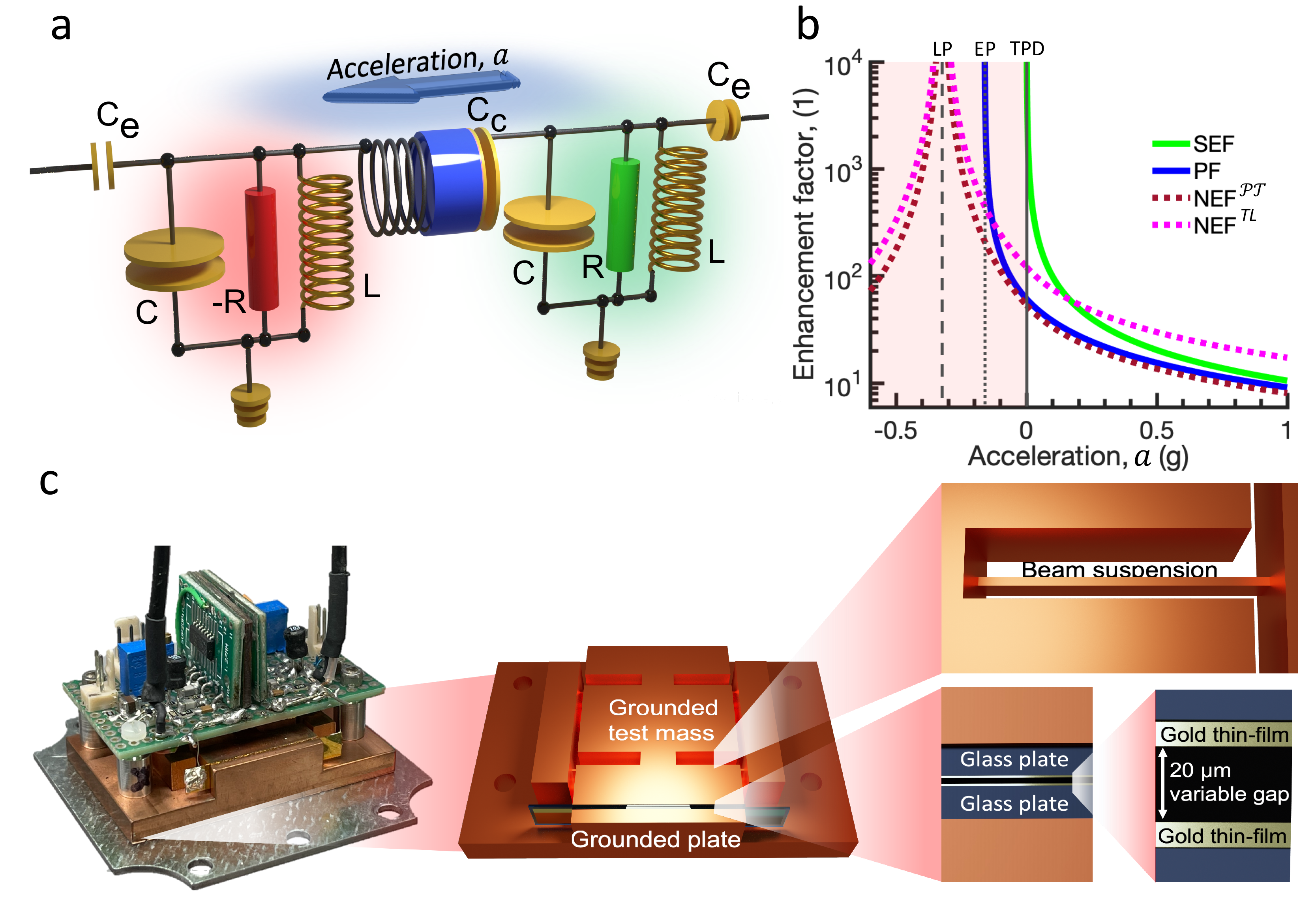}
\caption{\label{fig01} \linespread{0.5}\selectfont{} 
{\bf $\mathcal{PT}$ symmetric platform for enhanced acceleration sensing. } 
{\bf a,} Schematic of the $\mathcal{PT}$ symmetric electromechanical sensor. The ${\cal PT}$-symmetric circuit is capacitively 
coupled to the transmission lines with a capacitor $C_e$. The two circuit tanks are coupled together with a variable capacitor 
$C_c$ with one plate connected to a test mass that senses the acceleration. {\bf b,} The behavior of the sensitivity enhancement 
factor $SEF$ (green line), Petermann factor $PF$ (blue line), and of noise enhancement factors associated with the ${\cal PT}$-
components of the circuit $NEF^{\cal PT}$ (dark red dotted line) and the transmission line $NEF^{\rm TL}$ as a function of the 
applied acceleration $a$ (magenta dotted line). The black vertical lines indicate the TPD (solid), EP (dotted) and LP (dashed) points. The 
highlighted domain indicates the accelerations which are not captured by the experimental platform. {\bf c,} The actual acceleration 
sensing platform utilizes a micro-fabricated coupling capacitor which is connected to the test-mass used to sense the applied acceleration. 
}
\end{figure}


Here, we show experimentally and theoretically a ten-fold enhancement in responsivity to small perturbations and a three-fold 
signal-to-noise ratio (SNR) improvement in the sensing performance of an EP-based ${\cal PT}$-symmetric electro-mechanical 
accelerometer (see Fig. \ref{fig01}a). Our measuring protocol differs from the previous cases which were analyzing the lasing 
modes detuning of a system in the proximity to an EP. Instead, it relies on a distinction between EPs and the transmission peak 
degeneracies (TPD) observed in the transmission spectrum of the ${\cal PT}$-symmetric sensor when it is interrogated via 
weakly coupled transmission lines (TL). This fact has been already recognized in a previous theoretical work \cite{QGKDZ21} 
for the analogous concept of transmission deeps. Their distinct position from the position of the eigenmodes of the (isolated) 
system is traced to the interference effects between a direct scattering process and a resonant high-Q mode supported by 
the system. We point out that in the absence of loss/gain, these  transmission deeps (or peaks in our case) and their associated  degenerate point are related to the newly established concept of reflectionless scattering modes (RSM). The RSM frequencies 
belong to the spectrum of a wave operator with appropriate boundary conditions which has been shown to exhibit an EP 
degeneracy \cite{SHRS19,WRS20}. Nevertheless, the transmission deeps (peaks) and their associated TPD in the case of 
${\cal PT}$-symmetric systems are not related to the RSM exceptional points and the identification of an appropriate wave 
operator whose spectrum contains these modes is a topic of an ongoing research.

The transmission peaks frequency detuning from the TPD is influenced by the underlying EP and follows a square-root 
behavior with respect to the applied acceleration $a$. Furthermore, at TPD, the eigenbasis is still complete; thus the 
performance of the sensor is not influenced by excess noise effects that are rooted to the eigenbasis collapse. The latter 
has been related, in the platform of Ref. \cite{YWWGV20}, with the so-called Petermann factor ($PF$) which diverges in 
the proximity of an EP, while at TPD it remains finite and smaller than the sensitivity enhancement factor $SEF$ (compare 
blue and green lines in Fig. \ref{fig01}b) which measures the responsivity of the system to acceleration 
variations (see Supplementary Material). Our theoretical analysis indicates that $PF$ is directly proportional to the technical 
noise due to the coupling with the TL and the noise generated by the gain/loss elements used to create the EP-singularity. 
The corresponding noise enhancement factors $NEF^{TL}, NEF^{\cal PT}$, describing the noise power enhancements 
(see Supplementary Material), remain finite in the proximity of the TPD and are surpassed by the SEF of the transmission 
peaks detuning near the TPD (see brown and red lines in Fig.  \ref{fig01}b).

The sensing device consists of a pair of capacitively coupled parity-time ($\mathcal{PT}$) symmetric RLC resonators \cite{LRC} 
with natural frequency $f_0={\frac{1}{2\pi}}{\frac{1}{\sqrt{LC}}}\approx 2.68MHz$ and capacitive coupling $C_c$. Furthermore, one  
capacitor plate of the coupling capacitor is connected to a test-mass $m$ which is supported by a spring attached to the platform 
(see Fig. \ref{fig01}c and Methods for detailed platform design) and its displacement from an initial equilibrium position will constitute 
the mechanism that will sense acceleration variations (see discussion below). The $\mathcal{PT}$-symmetry condition is achieved 
when the gain (implemented using an amplifier) and loss parameters, namely $-R$ and $R$, are delicately balanced, and the 
reactive components, $L$ and $C$, satisfy mirror symmetry: that is, the impedances of the active and passive circuit tanks, multiplied 
by $i$, are complex 
conjugates of each other at the frequency of interest \cite{LRC}. When the capacitive coupling is above a critical value $C_c^{\rm EP}$, 
the system is in the so-called exact phase where the corresponding normal modes are also eigenvectors of the $\mathcal{PT}$
-symmetric operator. In the opposite limit (small coupling) the system is in the broken phase where the normal modes are no longer 
eigenmodes of the $\mathcal{PT}$ symmetric operator. In the latter case, the eigenfrequencies appear as a complex conjugate 
pair, while in the former domain they are real-valued. The two phases are separated by an EP singularity of order $N=2$ \cite{Bender}. 

\begin{figure}
\centering
\includegraphics[width=0.8\columnwidth]{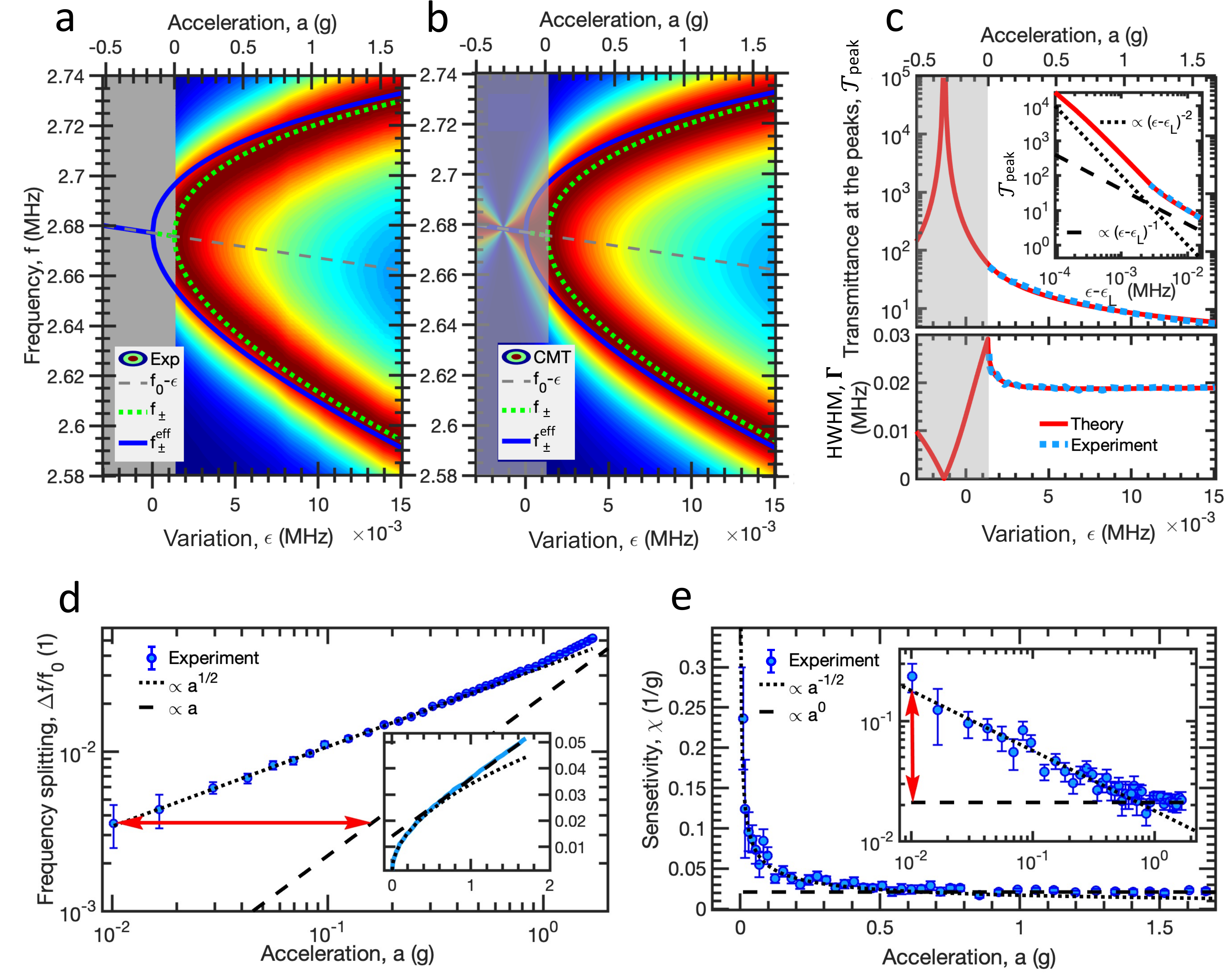}
\caption{\label{fig02} \linespread{0.5}\selectfont{} 
{\bf Experimentally measured response of the sensor to applied acceleration}. {\bf a,} Density plot of the measured normalized 
transmittance spectrum as a function of acceleration $a$ (top $x$-axis). The bottom $x$-axis indicates the associated capacitance 
variation $\epsilon(a)$. The trajectory of the transmittance peaks  is indicated with green dashed lines. At the proximity of the 
transmittance peak degeneracy (TPD) the peaks demonstrate a square-root behavior as expected by the Puiseux series expansion 
near EPs. The trajectories of the eigenfrequencies of the isolated dimer are shown by dark blue lines. They coalesce at $\epsilon
=0$ (EP degeneracy). The highlighted gray area is excluded from our measurements. In this domain, saturable nonlinearities are 
triggered. {\bf b,} The same as in {\bf a} for the transmittance spectrum calculated using a couped mode theory (CMT). {\bf c}, (Top) 
The transmission peak versus the capacitance variation  $\epsilon(a)$. The red line indicates the CMT predictions while the dotted 
light blue line is the measured data. The inset presents the same data in a double-logarithmic fashion referring to the capacitance 
variations with respect to the lasing point. The dashed and dotted black lines indicates a logarithmic slope of $-1$ and $-2$. (Bottom) 
The behavior of half-width-half-maximum linewidth $\Gamma$ versus $\epsilon$ in the proximity of the transmission peak degeneracy. 
The red (dotted light blue) line indicates the CMT (experimental) results.{\bf d}, The measured relative transmission peak splitting 
$\Delta f/f_0$ versus the applied differential acceleration $a$ plotted in double-logarithmic plot. The dotted (dashed) black line 
indicates a square-root (linear) scaling with $a$. Inset shows the same data in linear plot. The red arrow shows the dynamical 
range enhancement of the proposed accelerometer with respect to an equivalent sensor based on linear response. {\bf e}, The 
sensitivity of the ${\cal PT}$-symmetric accelerometer demonstrating an order (red arrow) enhancement in the proximity of the EP 
as opposed to a system configuration away from the EP. In the inset we report the same data in a double-logarithmic plot. Error 
bars on panels {\bf d} and {\bf e} denote $\pm$ 1 standard deviation obtained from 10 independent measurements.}
\end{figure}

We have {\it weakly} coupled each RLC resonator to $Z=50$ Ohm transmission lines (TL) via capacitors C$_e$. In the density 
plot shown in Fig. \ref{fig02}a, we report the measured (normalized) transmittance spectrum $T(f;a)$ versus the applied 
differential in-plane acceleration $a$. At $a=0$ the transmission spectrum demonstrates a transmission peak degeneracy (TPD) 
which reflects the nearby EP degeneracy of the eigenfrequencies of the isolated system (corresponding to $C_e=0$). The system 
is initially set at TPD conditions with the coupling capacitor plates being at a distance $d\approx 20\mu m$ from one another 
corresponding to $C_c^{\rm TPD}\approx 50pF$. As the applied differential in-plane acceleration $a$ increases, the capacitive 
coupling detunes from its equilibrium value $C_c^{\rm TPD}$ due to a displacement of the test-mass to a new equilibrium position. 
The working principle of our accelerometer relies on detecting concomitant transmission peak shifts from the TPD due to the 
capacitive coupling detuning from $C_c^{\rm TPD}$. Indeed, from Fig. \ref{fig02}a, we observe that the transmittance develops 
two equihight peaks which bifurcate from the TPD following a characteristic square-root behavior (see dashed green lines 
evaluated using a coupled mode theory Eq. (\ref{fpm})), expected from the Puiseux expansion of resonance frequencies around 
an EP (see blue lines). In fact, the responsivity at the TPD detuning, turns out to be slightly stronger from the analogues one 
shown by the resonances at the EP (see Eq. (\ref{approxf}) below). The weak coupling 
with the transmission lines, together with the active nature of our platform is reflected in the high intensity value of the transmittance 
peaks and in the narrow form of the linewidths (see confined dark red area). These attributes enhance the readout and boost the 
sensing resolution, allowing us to easily identify the trajectories of the  transmission peaks. Finally, the highlighted gray area indicates the 
displacement values of the capacitor plates for which the transmission peaks acquire extremely large values triggering saturable 
nonlinearities.

To better understand the outcome of these measurements we have performed a theoretical analysis using the most general 
framework of coupled mode theory (CMT) which was appropriately mapped to describe our $\mathcal{PT}$-symmetric circuit 
(see details in the Supplementary Material). Specifically, the scattering matrix $S$ that describes the open circuit, takes the 
form 
\begin{equation}
\mathbf{S}(f)=-\mathbf{I}-i\mathbf{W}\mathbf{G}(f)\mathbf{W}^T; \quad \quad\mathbf{G}(f)=(\mathbf{H}_{\rm eff}-f 
\mathbf{I})^{-1}
\label{Scat}
\end{equation} 
where, $\mathbf{I}$ is the two-dimensional identity matrix and $G(f)$ is the Green's function. The effective Hamiltonian 
$\mathbf{H}_{\rm eff}$ that describes the $\mathcal{PT}$-symmetric dimer coupled to the transmission lines is 
\begin{equation}
\begin{array}{ccc}
\mathbf{H}_{eff}=\mathbf{H}_{0}-\frac{i}{2}\mathbf{W}^{T}\mathbf{W};\quad
\mathbf{H}_0=\left( \begin{array}{ccc}
f_0-\epsilon+i \gamma_0 & \gamma_0+\epsilon \\
\gamma_0+\epsilon & f_0-\epsilon-i\gamma_0  \end{array} \right);\end{array}
\label{S3}
\end{equation} 
where  the diagonal matrix $\mathbf{W}_{nm}=\delta_{nm}\sqrt{2\gamma_e}$ models the coupling of the dimer with the 
transmission lines and $\mathbf{H}_{0}$ is the Hamiltonian of the isolated dimer (corresponding to $\gamma_e=0$). For 
$\epsilon=\epsilon_{\rm EP}=0$ the isolated system forms a second order ($N=2$) EP degeneracy at frequency $f_{\pm}^{(0)}
\equiv f_{\rm EP}=f_0$ (see Supplementary Material).

Using Eq. (\ref{Scat}) we have extracted the transmittance ${\mathcal T}(f;a)=|S_{21}|^2$ and via direct comparison with 
the experimental data we were able to identify the various parameters that have been used in our CMT modeling. 
Specifically, we have estimated that the linewidth broadening of the resonances of the individual circuit tank, due to 
its coupling with the transmission line is $\gamma_e\equiv Z_0\sqrt{C\over L} \left(C_e\over C\right)^2 f_0= 0.0206\,$MHz. 
Similarly, we have found that the variations at the coupling strength between the two resonant modes of the dimer, 
due to the displacement of the plates of the capacitor $C_c$ when an acceleration $a$ is imposed to the system, 
can be modeled by the parameter $\epsilon(a)=\epsilon_{\rm TPD}+0.0082$ [MHz/g]$\cdot a$. The coupling strength 
in the absence of any acceleration is $\epsilon_{\rm TPD}=0.0013$ MHz - and is associated with the TPD. Finally, $\gamma_0
= R^{-1} \sqrt{L/C} f_0=0.16$ MHz, is the CMT gain/loss parameter describing the amplifier/resistor used in the $\mathcal{PT}$
-symmetric circuit. 

Using the above extracted parameters we evaluated the transmittance spectrum ${\mathcal T}(f;\epsilon(a))=|S_{21}|^2$ 
from the CMT. The corresponding density plot is reported in Fig. \ref{fig02}b. The agreement between the theoretical and 
experimental results is reassuring for the quality of our fits and allow us to extract further information for the scattering 
characteristics of our system. Specifically, we have calculated the frequencies $f_{\pm}(\epsilon)$ of the transmittance 
peaks (green dashed lines in Figs. \ref{fig02}a,b) which take the form
\begin{equation}
f_{\pm}(\epsilon)=\left\{
\begin{array}{ccc}
f_0-\epsilon\pm\sqrt{2\gamma_0\epsilon+\epsilon^2-\gamma_e^2},& {\rm for} & \epsilon\geq\epsilon_{\rm TPD}\\
f_0-\epsilon, & {\rm for} & \epsilon\leq\epsilon_{\rm TPD}
\end{array}
\right.
\label{fpm}
\end{equation}
where $\epsilon_{\rm TPD} = -\gamma_0+\sqrt{\gamma_0^2+\gamma_e^2}\neq\epsilon_{\rm EP}$. Instead, we have that 
$\epsilon_{\rm TPD}\approx\frac{\gamma_e^2}{2\gamma_0}\xrightarrow[\gamma_e\rightarrow 0]{}\epsilon_{\rm EP}=0$ 
as expected.  Accordingly, the TPD frequency is $f_{\rm TPD}\equiv f(\epsilon=\epsilon_{\rm TPD})=f_-=f_+=f_0-\epsilon_{\rm 
TPD}\xrightarrow[\gamma_e\rightarrow 0]{} f_{\rm EP}=f_0$. Furthermore, at the vicinity of $\epsilon_{\rm TPD}$ the 
transmission peaks Eq. (\ref{fpm}) scale as 
\begin{equation}
f_{\pm}\propto f_{\rm TPD}\pm\sqrt{2}\sqrt[4]{\gamma_0^2+\gamma_e^2}\sqrt{\delta \epsilon};\quad {\rm provided}\,\,
{\rm that}\quad \delta\epsilon\ll 2\sqrt{\gamma_0^2+\gamma_e^2}
\label{approxf}
\end{equation}
where $\delta\epsilon\equiv \epsilon-\epsilon_{\rm TPD}$. It is important 
to stress that the above square-root transmission peak splitting for small $\delta \epsilon$ is a direct consequence of the 
presence of the EP degeneracies of the eigenfrequencies $f_{\pm}^{(0)}$ of the isolated system (e.g. $\gamma_e=0$). 
The latter, can be evaluated by direct diagonalization of $\mathbf{H}_{0}$, see Eq. (\ref{S3}) and their functional dependence 
on $\epsilon$ is given by Eqs. (\ref{fpm},\ref{approxf}) by substituting $\gamma_e=0$. It is interesting to point out that the 
presence of $\gamma_e$ results in a slight enhancement of the detuning rate of $f_{\pm}$ as compared to the detuning 
rate of the resonant modes $f_{\pm}^{(0)}$.

To guarantee the stability of our system, we have also identified theoretically the lasing condition of the open circuit (see 
Supplementary Material). The latter is associated with the real poles of the scattering matrix Eq. (\ref{Scat}) occurring at 
$f_L=f_0-\epsilon_L$ where $\epsilon_L=-\gamma_0+\sqrt{\gamma_0^2-\gamma_e^2}<\epsilon_{\rm EP}=0<\epsilon_{\rm 
TPD}$. Its existence can be also recognized by the divergence of the transmission at this frequency and perturbation values. 
In the upper subfigure of Fig. \ref{fig02}c we report the un-normalized transmission peaks ${\mathcal T}_{peak}={\mathcal T}
(f_{\pm};\epsilon)$ extracted from the measurements (dashed blue line), together with its theoretical values calculated using 
the CMT modeling (red line). We find that at the lasing point ${\mathcal T}_{peak}(\epsilon\rightarrow\epsilon_L)\rightarrow 
\infty$ as expected. The divergence is characterized by a ${\mathcal T}_{peak}(\epsilon)\propto (\epsilon-\epsilon_L)^{-2}
\quad  [{\mathcal T}_{peak}\propto(\epsilon-\epsilon_L)^{-1}]$ scaling, which applies in the range $\epsilon<\epsilon_{\rm TPD}
\quad [\epsilon>\epsilon_{\rm TPD}$ and it is nicely confirmed from the experimental data]. 

While an enhanced transmitted intensity is an important element for the readability of the output signal and the identification 
of the transmission resonances, the other crucial factor is the sensing resolution. This bound is proportional to the linewidth 
$\Gamma$ of the transmission peaks. In the limit of very strong coupling $\epsilon\gg \gamma_0$, we can 
disregard the $\mathcal{PT}$ nature of the circuit and we expect that the linewidth $\Gamma$ will be dictated by the coupling 
of the dimer with the transmission lines. The latter is characterized by the coupling constant $\gamma_e$ which in our setting 
takes the value $\gamma_e\approx 0.02\,$MHz. At the other limiting case of $\epsilon\rightarrow\epsilon_{L}$ we expect a 
narrowing of the linewidth which in the semiclassical picture (e.g. without taking into consideration the non-orthogonality of 
the modes etc) becomes zero. In fact, the CMT modeling provides us with the possibility to derive an exact expression for 
the linewidth $\Gamma(\epsilon)$ for the whole range of perturbations $\epsilon$ (see Supplementary Material). These CMT 
results are shown in the lower subfigure Fig. \ref{fig02}c together with the extracted values from the measurements which 
have been evaluated as the half-width at half-maximum (HWHM) at the transmission peak. Although $\Gamma(\epsilon)$ 
remains essentially constant $\Gamma(\epsilon)\approx \gamma_e$ at the parameter range where our experiment is 
operating, we are able to identify a slight increase in the proximity of $\epsilon_{\rm TPD}$. Specifically, our CMT analysis 
indicates that for $\delta\epsilon\equiv \epsilon-\epsilon_{\rm TPD}\geq 0$ the linewidth takes the form $\Gamma(\delta
\epsilon\rightarrow 0)\sim \sqrt{2}\gamma_e-\sqrt{2}\left(\gamma_e^2+\gamma_0^2\right)^{1/4}\sqrt{\delta\epsilon}$. For 
$\delta\epsilon=0$ we get that $\Gamma(\delta\epsilon=0)=\sqrt{2}\gamma_e$ indicating that the coupling of the circuit with 
the leads dictates the minimum measured uncertainty. This prediction is confirmed by the measurements (see light blue 
dotted line in Fig. \ref{fig02}c). On the other side of $\epsilon_{\rm TPD}$ i.e. $\delta\epsilon<0$ the linewidth behaves 
as $\Gamma(\epsilon)\approx \sqrt{2}\gamma_e+{\sqrt{2(\gamma_e^2+\gamma_0^2)}\over \gamma_e}\delta\epsilon$. This 
small $\delta\epsilon$ expansion persists even for $\epsilon\approx \epsilon_{\rm EP}$ where $\Gamma(\epsilon=\epsilon_{\rm 
EP})\approx 0.7\gamma_e$. From this analysis we conclude that the uncertainty in the frequency splitting measurements in 
the proximity of the TPD is only slightly enhanced due to the linewidth increase \cite{YSS11}. Nevertheless, it can be always 
confined to small values when the system operates in the weak coupling regime. Of course, this conclusion is subject to the 
analysis of other characteristics like the sensitivity or other noise sources that might also affect the measurement process.

The measurement of an individual frequency shift is the most common sensing scheme for (micro)-electromechanical sensors 
\cite{KWBLP12,13,15}. However, a considerable advantage is obtained by measuring the frequency splitting since it is 
intrinsically self-referenced i.e. there is no need for an external reference to suppress or eliminate frequency drift associated with 
other sources. From Eq. (\ref{approxf}) we have that the transmittance peaks splitting in the proximity of the TPD, also follows a 
sublinear square-root behavior with respect to the perturbation $\delta\epsilon$. Specifically $\Delta f \equiv f_{+}-f_{-}\propto 
\sqrt{\delta \epsilon}$. In case of weak coupling to the transmission lines the upper limit of the sublinear sensing is predominantly 
controlled by $\gamma_0$, see Eq. (\ref{approxf}). We point out that the extent of the sublinear domain in our 
system is bounded by nonlinear capacitance effects occurring at small distances between the plates of the capacitor. Even with 
this additional limitation our platform demonstrates an order of magnitude enhancement of dynamic range, as compared to 
conventional linear sensors with the same upper bound of dynamical range (see red double-side arrow in the inset of  Fig. 
\ref{fig02}d). An overview of the frequency splitting measurements are shown in Fig. \ref{fig02}d and nicely confirm the square 
root behavior expected from Eq. (\ref{approxf}). 

To further quantify the efficiency of our proposed accelerometer, we have introduced the sensitivity $\chi =\frac{\partial(\Delta
f/f_0)}{\partial a}$. In Fig. \ref{fig02}e, we report the experimentally measured sensitivity together with a theoretically derived 
curve (black dotted line). We find that indeed the sensitivity demonstrates an $\chi\sim 1/\sqrt{a}$ divergence in the proximity 
of the TPD. A comparison with the conventional linear sensor reveals again a sensitivity improvement of one order (red double-arrow).

\begin{figure}
\centering
\includegraphics[width=0.9\columnwidth]{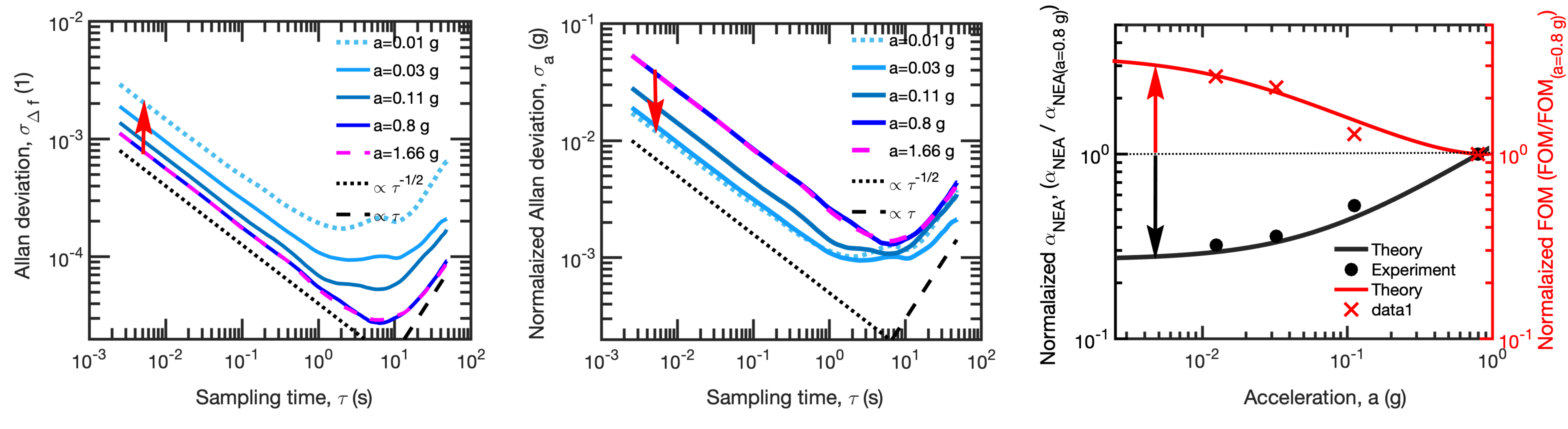}
\caption{\label{fig03} \linespread{0.5}\selectfont{} 
{\bf Measured Allan Deviation}. {\bf a,} Measured Allan deviation $\sigma_{\Delta f}$ of the transmission peaks splitting as 
a function of the sampling time $\tau$ for various values of the applied acceleration $a$ (see legend). The enhancement 
of $\sigma_{\Delta f}$ at the vicinity of the TPD at $a=0$ g is shown by red arrow. {\bf b,} Measured normalized Allan deviation $\sigma_{\alpha}=\sigma_{\Delta f}/\chi$ as a function of the sampling time $\tau$ for various values of the applied acceleration 
$a$ (see legend). The suppression of $\sigma_{\alpha}$ due to enhanced sensitivity at the vicinity of $a=0$ g is shown by 
red arrow. The black dotted lines on panels ({\bf a,b}) indicate a power decay $\tau^{-1/2}$, associated with a VRW regime, 
while the black dashed lines are proportional to $\tau$ indicating the presence of a RR regime. {\bf c} The normalized noise  equivalent acceleration $\alpha_{\rm NEA}/\alpha_{\rm NEA}(a=0.8$g) (black line and circles) and the normalized $FOM/
FOM(a=0.8$g) (red line and crosses) versus the applied differential acceleration. The black/red arrows indicate a threefold  reduction/enhancement of the noise/FOM in the proximity of the TPD. 
}
\end{figure}

{\it Noise Analysis -- } Enhanced eigenvalue splitting is an important aspect of efficient sensing as it guarantees an enhanced 
transduction coefficient of the sensor from the input quantity of interest (e.g. the differential acceleration) to the output quantity 
(e.g. $\Delta f$). Another important aspect of an efficient sensing is the precision of the measurement which is associated with 
the smallest measurable change of the input quantity given by the noise of the sensor output. A cumulative quantification of the 
noise effects on the measured frequency splitting $\Delta f$, is obtained by the analysis of the Allan deviation \cite{SHN08,QFFDF13} 
$\sigma_{\Delta f}$ for various values of applied differential in-plane acceleration $a$ (see Methods for details). The Allan deviation 
describes the stability of the sensor as a function of the sampling time $\tau$. From our analysis we find that $\sigma_{\Delta f}$, 
increases as the applied acceleration $a$ approaches zero, i.e. in the domain where the transmission peak splitting has a square
-root response. In contrast, we have found that at the domain where the frequency splitting exhibits a linear sensitivity ($a\geq 
0.5$ g) the Allan deviation $\sigma_{\Delta f}$ remains almost unchanged (compare dark blue solid line and magenta dash line 
in Fig. \ref{fig03}a). Such behavior, indicates that the noise is enhanced at the vicinity of $a=0$ g (where $\epsilon=\epsilon_{TPD}$) 
(see red arrow in Fig. \ref{fig03}a).

A more appropriate characterization of the performance of our sensor requires the analysis of the normalized Allan Deviation 
$\sigma_\alpha=\sigma_{\Delta f}/\chi$. Such analysis provides an estimate of the total noise effect on the measured acceleration. 
In particular, the short
-time decay of the normalized Allan deviation (Velocity Random Walk) behaves like $\sigma_{\rm VRW}(\tau)=\alpha_{\rm NEA}
\cdot\tau^{-1/2}$, with a slope $\alpha_{\rm NEA}$ which decreases from $\alpha_{\rm NEA}=0.0027$g$\cdot$ Hz$^{-1/2}$ at  $a=1.66$g to a limiting value $\alpha_{\rm NEA}=0.00086$g$\cdot$ Hz$^{-1/2}$, at $a=0.01$g. The experimental data shown 
in Fig. \ref{fig03}b indicates that there is a three-fold SNR improvement at the vicinity of TPD. 

The noise-equivalent acceleration (NEA) $\alpha_{\rm NEA}$ can be further expressed as a sum of various terms 
associated with a different noise sources that might affect the precision of the measurements. Specifically we have:
\begin{equation}
\alpha_{\rm NEA}^2(\epsilon)\equiv \alpha_{\rm th}^2+\alpha_{\rm TL}^2+
\alpha_{\mathcal{PT}}^2+\alpha_{\rm add}^2,\quad \left[{\rm units}\,\, {\rm of}\, {\rm g}\cdot{\rm Hz}^{-1/2}\right].
\label{E2}
\end{equation}
The first term arises from a thermal Brownian motion of the test mass and is given by $\alpha_{\rm th}=\sqrt{4k_{\textrm{B}} 
T \omega_n/mQ}$. In this expression $k_{\textrm{B}}$ is the Boltzmann's constant, $T=293$ K is the ambient temperature, 
$m=27$gr is the mass, $\omega_n=2\pi f_n= 2513\,$ rad/sec is the frequency and $Q=40$ is the quality factor of the resonant 
mode of the mechanical test-mass. The reduction of the {\it thermal noise equivalent acceleration} is fundamental in minimizing 
the NEA towards its standard quantum limit corresponding to the position uncertainty in the quantum ground state of the mechanical 
mass. Typically it is overrun by the other terms in Eq. (\ref{E2}). The other two terms arise from the thermal noise of the transmission 
lines ($\alpha_{\rm TL}$), the $\mathcal{PT}$-circuit noise ($\alpha_{\mathcal{PT}}$) produced by the gain-loss elements of the 
platform. The remaining term is the added noise ($\alpha_{\rm add}$) associated with the fluctuations due to the Brownian motion 
of the plates of the coupling capacitors $C_{c2}$ and $C_{v}$ (see Methods) and the thermal noise produced from the movement 
of various mechanical parts of the experimental setup  (see Methods). Both $\alpha_{\rm th}$ and $\alpha_{\rm add}$ are not 
associated with (and do not depend on) parameters of the $\mathcal{PT}$ symmetric circuit itself and thus, do not depend on $\epsilon$.


One can further understand the demonstrated sensing efficiency by realizing that $\alpha_{\mathcal{PT}}\equiv 
{\widetilde \sigma}_{\mathcal{PT}}/\chi$ and $\alpha_{\rm TL}\equiv{\widetilde \sigma}_{\rm TL}/\chi$ where 
${\widetilde \sigma}_{\mathcal{PT}}$ and ${\widetilde \sigma}_{\rm TL}$ are the spectral densities of variances in 
the acceleration measurements due to noise sources associated with the amplification/attenuation elements of 
the isolated circuit and with the transmission lines respectively. These spectral densities are proportional to the 
resonance linewidth $\Gamma$ and to the noise enhancement factors associated with the corresponding noise 
sources while they are inverse proportional to the transmitted signal through the electronic circuit (see Supplementary 
Material). The theoretical analysis based on CMT indicates that both ${\widetilde  \sigma}_{\mathcal{PT}}$ and 
${\widetilde \sigma}_{\rm TL}$ do not experience strong variations with respect to $\epsilon$ and both saturate 
to a finite value at $\epsilon=\epsilon_{\rm TPD}$. Since $\chi$ diverges at $\epsilon_{\rm TPD}$ (see green line 
in Fig. \ref{fig01}b), the NEA will decreases as $\epsilon$ approaches $\epsilon_{\rm TPD}$ towards its noise floor 
level $\alpha_{NEA}^2(\epsilon\rightarrow \epsilon_{TP}) \rightarrow \alpha_{th}^2 + \alpha_{\rm add}^2$. 
To appreciate further the enhancement of SNR in the proximity of the $\epsilon_{\rm TPD}$ we have normalized the 
NEA with respect to its value at large accelerations e.g. $a=0.8\,$g. The normalized data (black circles) are shown 
in Fig. \ref{fig03}c and demonstrate clearly a three-fold noise reduction in the proximity of TPD. The CMT results 
are also indicated in this figure with a black solid line and nicely match the experimental findings. Furthermore, the 
CMT analysis shows that the noise associated with the TL coupling and gain/loss elements used to create the sensor 
singularity are equivalent to voltage (or current) noise sources in the circuit. These are physically distinct from the 
parametrically coupled signal transduction sensing mechanism, and are transformed differently by the transmission 
singularity, toward a favorable signal-to-noise ratio.

We conclude that in the specific case of accelerometers the lower limit of NEA is dictated by the Brownian motion 
of the test mass and the capacitor plates which scales proportionally to the sensitivity $\chi$. We stress that on the modeling 
level, the Brownian motion has to be modeled as a fluctuation of the Hamiltonian matrix elements (i.e. coupling coefficient) 
while the electronic thermal noise due to amplifiers, resistors and transmission lines is described by additive stochastic 
Langevin terms in the Hamiltonian matrix. The electronic thermal noise, does not scale with 
$\chi$; it rather abides to a noise-specific transfer function which in the case of the PT-symmetric circuit is proportional 
to the $PF$. Therefore, whenever the sensitivity $\chi$ overwhelms the noise-specific transfer function, the limiting 
value of $\alpha_{\rm NEA}$ is the same for any sensing (linear or sub-linear) scheme. An example of a linear sensing 
protocol that demonstrates extreme sensitivity is associated with slow light 
sensors, utilizing high-Q resonant modes like the ones occurring at the band-edges of a Fiber Bragg grating (FBG) \cite{FBG1,
SKOLIANOS2}. Such sensors rely on the abrupt intensity variations of the transmission of a CW when a perturbation (e.g. stress) 
modifies the effective refractive index of the FBG, thus inducing a spectral shift of the high-Q Lorentzian resonance. This enhanced 
sensitivity is offset by an extremely short dynamic range that these sensors demonstrate: while the large sensitivity $\chi$ leads to a 
reduction of the lower sensing bound by minimizing the NEA, it simultaneously lowers the upper sensing bound, which is typically 
limited by the maximum measured signal that the sensor can output. In contrast, the enhanced sensitivity demonstrated by the 
TPD sensing scheme reduces, from one hand the lower sensing bound, while from the other hand, is not affecting the upper sensing 
bound of the measured accelerations. Additionally, in linear sensors the high sensitivity is inversely proportional to the mechanical 
resonant frequency of the test mass $f_n$. The latter dictates the operational bandwidth of the sensor - if the frequency of the applied 
acceleration signal exceeds the $f_n$ this signal will not be detected by the sensor. As a result, the high sensitivity of a linear sensor 
also comes at a cost of its bandwidth. Based on these key figure of merits, the performance of our proposed TPD-based proof-of-concept 
accelerometer compares favorably with the available on-market on-chip accelerometers such as NXP MMA1270KEG \cite{ACCEL}. 
For example, they both demonstrate an $\alpha_{\rm NEA}=0.00086$g$\cdot$ Hz$^{-1/2}$, for a measurement range up to ($\geq 2$) 
g, while our prototype has a 6.5 times larger operational bandwidth.

An alternative way to reach the ultimate bound of the SNR of a linear sensor is by increasing the power of the input signal to ``infinite" 
levels. Of course, this is practically challenging – if not impossible. In our case, a finite power consumption enhancement is automatically 
taking place when the system operates at the proximity of the TPD. It is therefore imperative to take also this effect into consideration when 
comparing the efficiency of a sensor. This point has been recently raised by Alu and coworkers which have introduced an additional Figure 
of Merit (FOM) that takes into consideration the consumed power $U$ used for the sensing \cite{DMA21}. We have expressed FOM in terms 
of $\alpha_{\rm NEA}$ as $FOM=(\alpha_{\rm NEA}^2P)^{-1}$ and plotted in Fig. \ref{fig03}c the rescaled FOM (with the value $FOM(a=
0.8\,$g) extracted from the CMT (red solid line) and the experiment (red crosses) using the measured $\alpha_{\rm NEA}$. The analysis 
indicates an improved FOM of the ${\cal PT}$-symmetric sensor in the proximity of TPD.

Let us also discuss the opposite limit of long-time growth of $\sigma_{\alpha}(\tau)$. This analysis allows us to extract the 
so-called drift rate ramp (DRR), describing systematic (deterministic) errors due to temperature fluctuations. By definition  $\sigma_{DRR}(\tau)=\alpha_{\rm DRR}\cdot\tau$, with best fit $\alpha_{\rm DRR}\approx 0.0007$ g$\cdot$s$^{-1}$ for 
almost all acceleration values. Finally, the saturation value $\sigma_{BI}(\tau) = \alpha_{\rm BI} \cdot\tau^0$ of the Allan 
deviation (black dashed line) is indicative of the Bias Instability (BI), and sets the smallest possible reading of our sensor 
due to the random flickering of electronics or other components. The extracted value ranges from $\alpha_{BI}\approx 
0.0009$ g (accelerations closer to TPD) to $\alpha_{BI}\approx 0.0014$ g (accelerations away from TPD).

{\it Conclusions --} We have studied theoretically and experimentally the sensing performance of a ${\cal PT}$-symmetric electromechanical accelerometer in the proximity to an exceptional point (EP) degeneracy. When the system is weakly 
coupled to transmission lines, the spectrum demonstrates transmission peak degeneracies (TPDs) which are influenced 
by the presence of the EP but occur at distinct parameter values. Our findings indicate that the TPDs inherit the enhanced 
sensitivity of the EP which, however, is not offset by the slight noise enhancement. We measured a three-fold signal-to-noise 
ratio enhancement at the vicinity of TPD. Although our analysis was focusing on avionic sensing, our platform can also be 
used for supersensitive microfluid flow sensors, pressure sensors or for the detection of magnetic field variations. Our results, 
show promise for the use of EP-based platforms as a novel scheme for enhanced sensing without the concerns associated 
with the noise amplification. 


\setcounter{figure}{0}
\renewcommand{\thefigure}{Extended Data Figure \arabic{figure}}

\section{Methods}
\subsection{Transmittance spectrum and frequency splitting measurements}
The transmitted signal through the electronic circuit was collected for different applied in-plane projections of gravity acceleration. 
The device was mounted on a rotational stage Newport ESP100BCC which was electronically controlled via a Newport ESP301 
controller. The setup allows to control the horizontal tilt of the device with minimal increment of 0.2 mdeg and, thus, vary the in-plane 
projection of the gravity acceleration in a range of 2g ([-1 g :1 g]). At each specific tilting angle the transmission spectrum was 
collected using an ENA network analyzer Keysight E5080A. The individual frequency sweeps contain 401 points in a range of 2.58 
MHz - 2.74 MHz, with individual frequency bandwidth of 1.5 MHz. Single measurement is obtained from the collected spectrum which was averaged over 100 consecutive 
individual sweeps for each applied acceleration, which resulted in a sampling time of $\tau\approx 25$ ms. The frequencies of the 
transmittance peaks $f_{\pm}$ were then identified from the resulted spectrum, which allows to calculate the frequency splitting 
$\Delta f$.

\subsection{Allan deviation measurement}
The Allan deviation $\sigma_{\Delta f}(\tau)$ of the transmission peak splitting is defined as
\begin{equation}
\label{Allan}
\sigma_{\Delta f}(\tau)=\sqrt{{1\over 2(M-1)} \sum_{n=1}^{M-1} (\overline{\Delta f}_{n+1}-\overline{\Delta f}_n)^2}
\end{equation}
where $\tau$ is the sampling time, $M$ is the total number of frequency measurements, and $\overline{\Delta f}_n$ indicates the 
average frequency splitting during the sampling time interval $[n\tau,(n+1)\tau]$. For the extraction of Allan deviation the transmittance 
peak splittings $\Delta f$ were sampled with frequency of 400 Hz ($\tau_{min} = 2.5$ ms) over a period of 90 s.

\subsection{Figure of Merit (FOM) and consumed power}

From the CMT analysis (see Supplementary material) one can show that:
\begin{equation}
\label{Energy}
|S_{21}|^2+|S_{11}|^2+2\gamma_2|a_2|^2-2\gamma_1|a_1|^2-1=0; \quad {\bf a}=-i{\bf GW}^T(1,0)^T,
\end{equation}
where the $\gamma_1/\gamma_2$ is the gain/loss of the first/second cavity. In our settings $\gamma_1=\gamma_2=\gamma_0$. The total power $P$ consumed by the platform is
\begin{equation}
\label{Energy2}
P= \int_{f_{min}}^{f_{max}} \left(|S_{21}|^2+|S_{11}|^2+2\gamma_2|a_2|^2\right)|\mathbf{a}_{{\rm TL}_1}^{in}|^2\,d f= \int_{f_{min}}^{f_{max}} \left(2\gamma_1|a_1|^2+1\right)|\mathbf{a}_{{\rm TL}_1}^{in}|^2 df.
\end{equation}
In our experimental platfrom  $f_{min}=2.58$ MHz and $f_{max}=2.74$ MHz, which are the lower/upper bound of frequency range used to perform the measurements.  Taking into account the fact that the input signal $\mathbf{a}_{{\rm TL}_1}^{in}$ has a uniform spectral density, the $FOM$ introduced in Ref.\cite{DMA21} can be written as
\begin{equation}
\label{FOM}
FOM=(\alpha^2_{NEA}P)^{-1} \propto\int_{f_{min}}^{f_{max}} (\alpha_{NEA}\left(|S_{21}|^2+|S_{11}|^2+2\gamma_2|a_2|^2\right))^{-1}\, df.
\end{equation}
We used the $S_{21}$ measurements to fit the measurements with CMT  parameters,  allowing us to estimate the total power $P$ using Eq.\ref{Energy2}. Finally, along with experimental data of $\alpha_{NEA}$ this allows to estimate the $FOM$ using Eq.\ref{FOM}.

\subsection{Circuit design and fabrication}n
The detailed schematics of the ${\cal PT}$-symmetric circuit is shown in the Extended data Fig. \ref{FIGE01}. The main elements of 
the circuit are a pair of RLC resonators, where resistor, inductor and capacitor are connected in parallel to ground. The inductor L in 
each unit is a Murata 11R103C with inductance of $L=10$ $\mu$H. The total capacitance consists of a pair of in-parallel connected 
capacitors C and C$_v$, where the former one is Murata GRM21A5CE271JW01D with fixed value of $C=270$ pF, while the latter 
one is a voltage controlled capacitor Murata LXRW0YV33-0-056 with variable capacitance $C_v=16.5-33$ pF. It is used to precisely 
control the resonant frequency of each resonator. External voltage was used in order to control the $C_v$. It was applied via an 
EG$\&$G 7265 DSP Lock-in amplifier, connected via a BNC port through a resistor R$_v$ Yageo RC0402FR-75KL with resistance 
$R_{v}=5$ kOhm and grounded $10$ $\mu$F capacitance C$_{v1}$ Murata GCM32EL8EH106KA7. The total resistance in each 
resonator unit consists of a fixed resistor R$_1$ Bourns CR1206FX-1101ELF with resistance $R_1=1.1$ kOhm, which is connected 
in-series with a mechanically controlled variable resistor R$_2$ Bourns 3296'w-1-20RLF with resistance $R_2=0-200$ Ohm. 

In the first resonator these resistors are connected to the output voltage port of an operational amplifier (Op Amp) in order to produce 
gain. The Op Amp used in the circuit is one of the three on-chip Op Amps of the Analog Devices ADA4862-3 triple amplifier. The internal 
resistance between the inverting port to ground is $R_{G1} = 550$ Ohm . This is the same as the resistance between the output port 
and the inverting input port $R_{G2} = 550$ Ohm (see extended data Fig. \ref{FIGE01}). The non-inverting input of the amplifier unit 
is connected to the node of the first resonator via the resistor Rs, Bourns RC0402FR-075KL with resistance $R_s=133$ Ohm. The 
second unit of the same ADA4862-3 triple amplifier is connected with non-inverting input to the node of the first resonator through the 
same resistance $R_s=133$ Ohm. The resistor $R_{G2}$ is left floating in order to obtain a unity gain. The tripple amplifier is powered 
via the voltage source through ports -Vs and +Vs, as shown in extended data Fig. \ref{FIGE01}. The output port of the Op Amp unit is 
connected with the probing input BNC port via the capacitor Ce, Murata GCM1887U1H103JA6J with capacitance $C_e=10$ nF. This 
capacitance was used to block the DC-signals from entering the circuit, while the coupling of the input probing signal is achieved via 
parasitic capacitive coupling between the input BNC port and the circuit elements. The resistors R$_{e1}$, Bourns CRM1206FX2700ELF 
with resistance $R_{e1}=270$ Ohm and R$_{e2}$, Bourns CR1206FX60R4ELF with resistnace $R_{e1}=50$ Ohm are connected as 
shown in the extended data Fig. \ref{FIGE01} in order to match the circuit impedance with the 50 Ohm coaxial cable connected to the 
BNC port.

In the second resonator unit the resistors R$_1$ and R$_2$ are connected to ground in-parallel to the capacitors C and C$_v$. The 
probing output BNC port is connected to the node of the second resonator through the Op Amp unit of another ADA4862-3 triple 
amplifier. The Op Amp unit operates in a unity gain regime (similar to the second Op Amp connected to the node of the first resonator).

The two resonator units are capacitevely coupled together using a trimmed capacitor C$_{c2}$ Knowles JZ300HV with variable 
capacitance $C_{c2}=5.5-30$ pF. The spring-mass is coupled to the circuit producing an extra coupling capacitance $C_{cv} \approx 35$ pF, 
when the spring-mass is at rest. When an in-plane acceleration is applied, the mass with the attached capacitor plate moves toward 
the stationary plate, resulting in an increase of the capacitance $C_{cv}$. The circuit was built on a prototype board and was mounted 
on top of the spring-mass as it is seen in Fig. \ref{fig01}. The electrodes of the spring mass were connected to the circuit with 0.18 mm 
wires which provides enough flexibility during the motion of the test mass. The wire were glued to the electrodes using silver filled epoxy 
adhesive MG Chemicals 8331-14G.

\subsection{Spring-mass capacitive sensor design and fabrication}
\begin{figure}
\centering
\includegraphics[width=1\columnwidth]{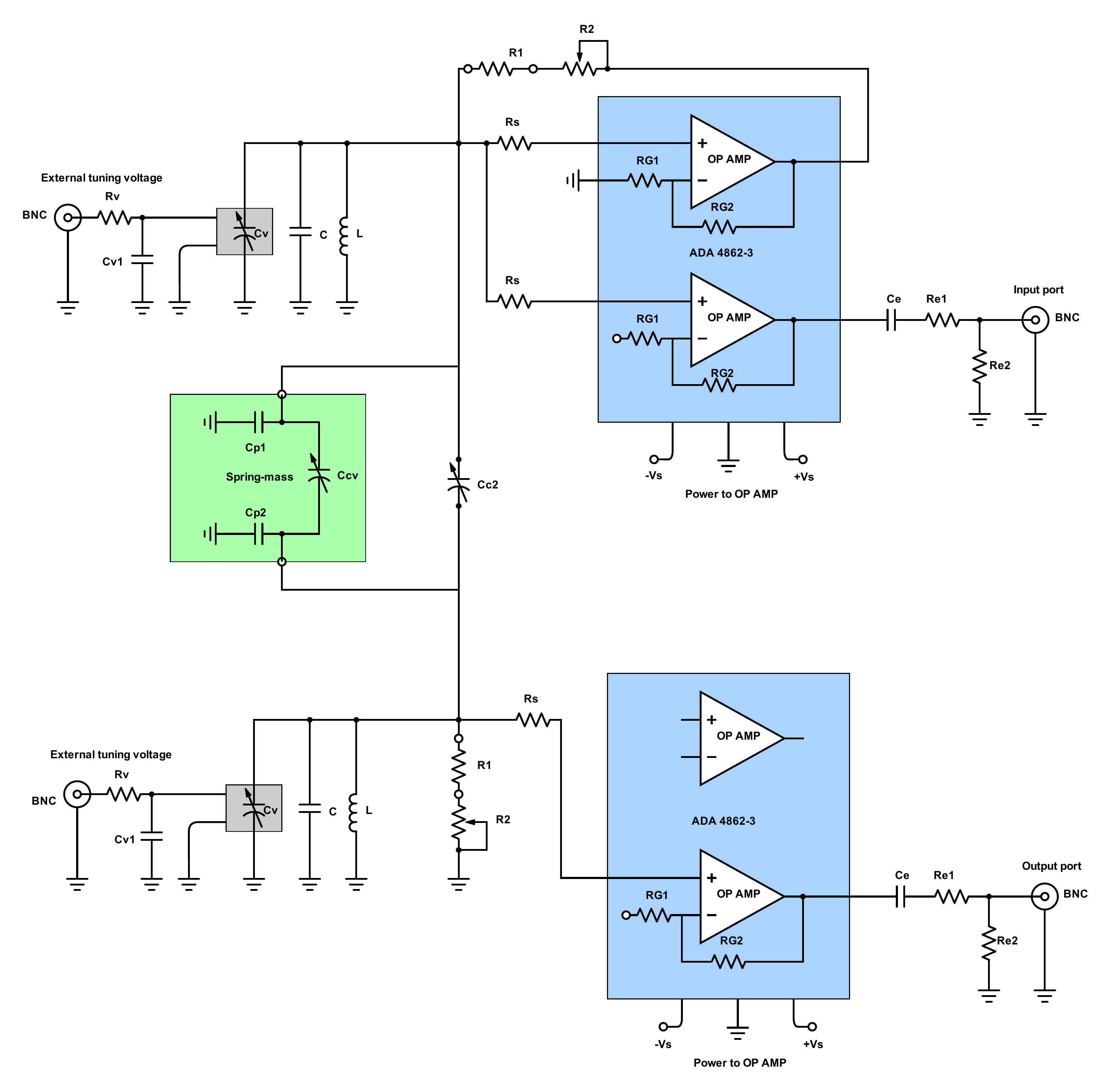}
\captionsetup{labelformat=empty}
\caption{\label{FIGE01} \linespread{0.5}\selectfont{} 
{\bf Extended Data Figure 1:}\, Schematic of the circuit diagram. The blue elements are  ADA4862-3 amplifiers, while the green element indicates 
the spring-mass which provides the acceleration dependent capacitance C$_{cv}$. }
\end{figure}

\begin{figure}
\centering
\includegraphics[width=1\columnwidth]{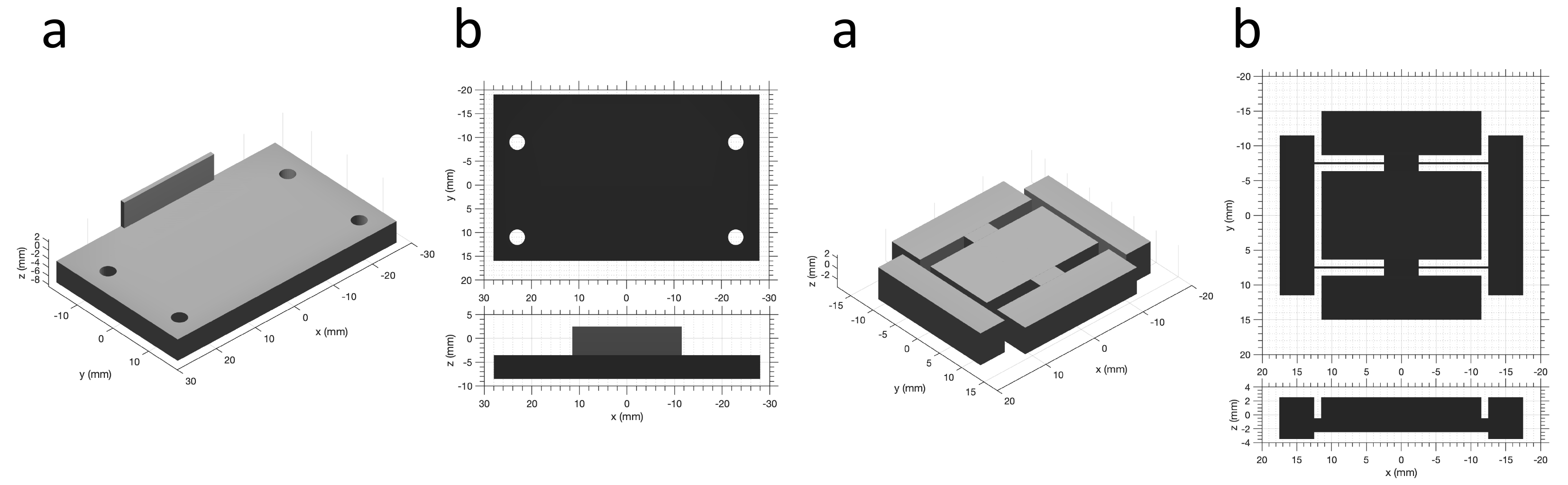}
\captionsetup{labelformat=empty}
\caption{\label{FIGE02}
{\bf Extended Data Figure 2:}\,
Details of the mechanical sensor elements. {\bf a,b}, Drawings of the copper platform. {\bf c,d}, Drawings 
of the test-mass. }
\end{figure}

\begin{figure*}
\centering
\includegraphics[width=1\columnwidth]{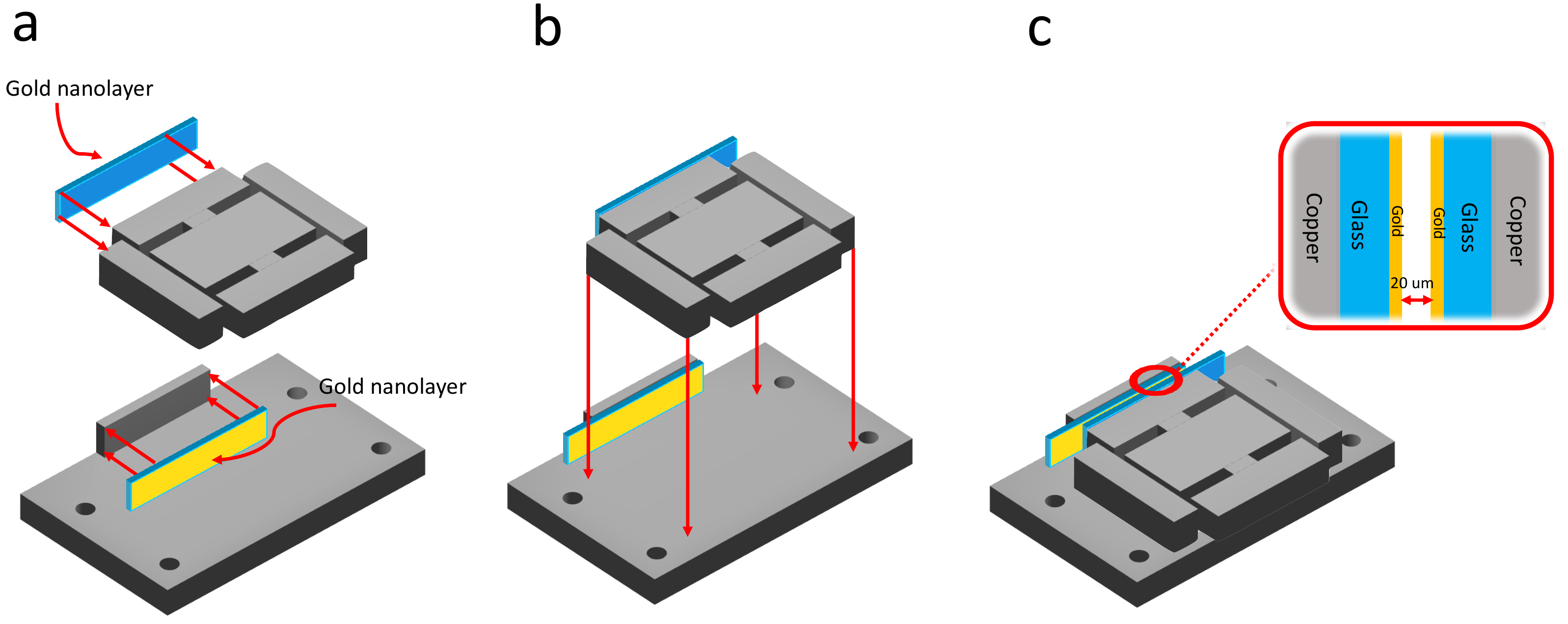}
\captionsetup{labelformat=empty}
\caption{{\bf Extended Data Figure 3:}\, \linespread{0.5}\selectfont{} 
Assembly process of the acceleration capacitive sensor. {\bf a}, Deposition of the gold nanofilms on a glass 
substrate that create conductive electrodes which form the capacitors plates. Attachment of the glass plates to the stationary platform 
and test-mass. {\bf b}, Placement of the test mass with glass plate on top of the copper base. {\bf c}, Assembled capacitive inertial sensor. 
The inset shows the magnified view of the area between the capacitor plates which is about 20 um.     }
\label{FIGE03}
\end{figure*}

The spring-mass capacitive sensor consists of two main pieces. The first one is a base plate shown in Figs. \ref{FIGE02}a and b. It 
consists of a 56 x 25 mm x 5 mm copper plate with one side having a 20 mm long ledge with 6 mm height. The ledge is used to mount 
the stationary capacitor plate. The second main piece is a 0.027 kg, 5 mm thick copper mass which is suspended on four 10 x 0.3 x 2 
mm bendable beams attached to stationary stands (see Figs. \ref{FIGE02}c and d). The mass is suspended 1 mm above the bottom 
level of the stationary stands. The mass and base plates were micro machined using electrical discharge machining (EDM) technique. 
The pair of 23 x 0.5 x 5 mm glass substrates were used to sputter (PVD) a 10 nm thin Ti layer followed by a 100 nm thin Au layer which 
acts as electrodes for the capacitor plates. One of the glass plates was glued to the ledge of the base plate (as it is shown in Fig. 
\ref{FIGE03}a) at the noncoated side. In a similar manner the second plate was glued to the back vertical surface of the copper mass 
at the noncoated side (see Fig.\ref{FIGE03}b). The copper mass with the glass plate was then mounted on the copper base and glued 
with a conductive adhesive (see Fig. \ref{FIGE03}c). The assembly was performed under a microscope with a custom built micro-translational 
stage to achieve the 20$\mu$m gap between the two gold-metalized surfaces of the glass plates (see inset
in Fig. \ref{FIGE03}c). The spacing between the pair of glass plates with gold nanolayers forms a capacitance of about 35 pF, when 
the mass is at rest. When an acceleration is applied to the platform, the plates are moving closer to one another, leading to an increase 
of the capacitance. The open parts of the metalized surfaces were connected to the circuit as it is discusses above, while the platform 
and the mass are grounded. The pair of glass plates form the parasitic capacitances $C_{p1}$ and  $C_{p2}$ between the gold electrodes 
and the grounded test mass. This parasitic capacitances are shown in Fig. \ref{FIGE01} and originally result in a frequency missmach 
between the LC resonators. The tuning capacitors C$_v$ were used to compensate the associated frequency mismatch between the 
resonators.   

\hspace*{-1cm}\textbf{Acknowledgments: } We acknowledge partial support from NSF-CMMI-1925543, NSF-CMMI-1925530, ONR N00014-19-1-2480 and from a grant 
from Simons Foundation for Collaboration in MPS No. 733698. RT and JC also acknowledge 
the partial support for this research provided by the University of Wisconsin-Madison, Office of the Vice Chancellor for Research and Graduate Education 
with funding from the Wisconsin Alumni Research Foundation.\\[-1.5cm]

\hspace*{-1cm}\textbf{Author contributions: } R.K., J.C., F. E. and R.T. designed the mechanical device. R.K. and F. E. designed and fabricated the electronic 
circuit. J.C. and R.T. fabricated the mechanical device. R.K. performed the characterization and data processing of the accelerometer and 
developed the theory with the support of T.K.  All authors discussed the results. T.K. conceived the project. R.K. and T.K. wrote the manuscript 
with input from all authors.\\[-1.5cm]
 
\hspace*{-1cm}\textbf{Competing interests: }The authors declare that they have no competing interests.\\[-1.5cm]

\hspace*{-1cm}\textbf{Data availability :} Data are available from the corresponding author upon reasonable request.\\[-1.5cm]

\hspace*{-1cm}\textbf{Code availability :} All codes are available from the corresponding author upon reasonable request.

\newpage

\pagebreak

\begin{center}
\textbf{\large Supplementary Material}
\end{center}

\setcounter{equation}{0}
\setcounter{section}{0}
\setcounter{figure}{0}
\setcounter{table}{0}
\setcounter{page}{1}
\makeatletter

\renewcommand{\thesection}{S\arabic{section}}  
\renewcommand{\thefigure}{S\arabic{figure}}
\renewcommand{\theequation}{S\arabic{equation}}

\section{CMT modeling of the electronic circuit}

\begin{figure}
\centering
\includegraphics[width=0.6\columnwidth]{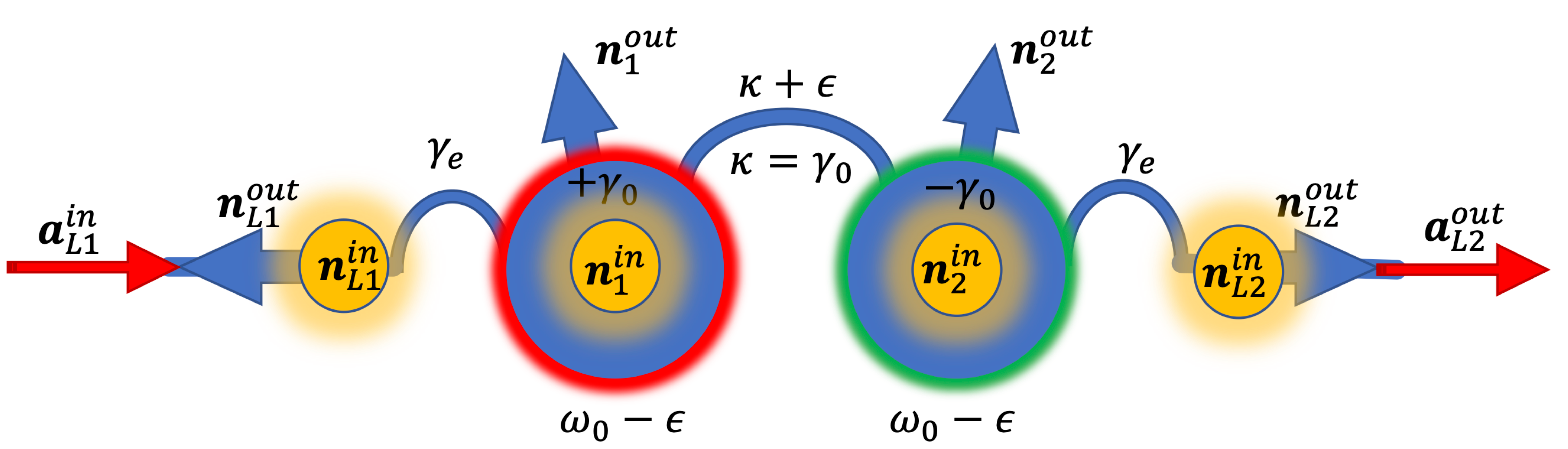}
\caption{ \linespread{0.5}\selectfont{} 
{\bf Schematic of the coupled mode theory setup: } The platform consists of two resonators with resonant frequency 
$\omega_0$ and balanced gain $+\gamma_0$ (red circle), and loss $-\gamma_0$ (green circle) elements. The system is 
prepared in a way that the coupling strength between the two resonators is $\kappa=\kappa_{\rm EP}=\gamma_0$. The variations in the 
coupling (due to e.g. differential acceleration) are probed by sending an input signal $\mathbf{a}_{{\rm TL}_1}^{in}$ from the 
left transmission line and measuring the outgoing signal $\mathbf{a}_{{\rm TL}_2}^{out}$ at right transmission line. The coupling strength 
between the coupled mode system and the transmission lines is denoted by $\gamma_{e}$. The resonant modes are 
affected by an ambient noise associated with the right and the left transmission lines $\mathbf{n}_{{\rm TL}_1/{\rm TL}_2}^{in}$ and 
by the noise $\mathbf{n}_{1,2}^{in}$ generated internally due to the presence of the gain (left) and the loss (right) mechanisms. 
The system emits noise to the left/right transmission lines $\mathbf{n}_{{\rm TL}_1/{\rm TL}_2}^{out}$ and to the left/right internal reservoirs 
$\mathbf{n}_{1/2}^{out}$. When the system experiences an external perturbation (e.g. acceleration) the coupling constant
between the two resonators is modified as $\kappa \rightarrow \kappa_{\rm EP}+\epsilon$; thus producing a detuning of the resonant 
frequencies $\omega_0\rightarrow \omega_0-\epsilon$ of the individual resonators.
}
\label{figs1}
\end{figure}

The electronic circuit of Fig. 1a can be modeled using a CMT picture, see Fig. \ref{figs1}. This mapping allows us to 
extend the conclusions of our investigations to a broader family of systems where CMT is applicable. At the same 
time it provides a simple mathematical description of our experimental set-up.

Let us first start from the description of an $LC$ circuit, which is described by a set of equations:
 \begin{equation}
\centering
\Bigg\{ \begin{array}{ccc}
  \mathcal{V} = L \frac{d\mathcal{I}}{dt} \\
  \mathcal{I} = - C\frac{d\mathcal{V}}{dt}.
  \end{array}
  \label{CMT1}
\end{equation}
The above equations can be re-written in the form of a second order differential equation for the 
voltage:  
\begin{equation}
\mathcal{\ddot{V}}+\omega_0^2\mathcal{V}=0.
\label{CMT3}
\end{equation} 
which can be solved easily, giving 
\begin{equation}
\centering
\Bigg\{ \begin{array}{ccc}
  \mathcal{V} =|V|cos(\omega_0t+\phi) \\
  \mathcal{I} = \sqrt{\frac{C}{L}}|V|sin(\omega_0t+\phi),
  \end{array}
  \label{CMT2}
\end{equation}
where $|V|$ is the voltage amplitude of the LC resonator, $\phi$ is the associated phase and $\omega_0 = 
\frac{1}{\sqrt{LC}}$ is the natural frequency of the circuit.  

It is useful to change variables and define the complex mode amplitude $a(t)$ and its conjugate $a^*(t)$  
requiring the normalization: $|a(t)|^2 = W = \frac{C}{2}|V|^2$, where $W$ is an energy stored in the  circuit. 
This yields that 
\begin{equation}
\centering
 \begin{array}{ccc}
  a = \sqrt{\frac{C}{2}}(\mathcal{V} -i\sqrt{\frac{L}{C}}\mathcal{I}) = \sqrt{\frac{C}{2}}(\mathcal{V} -
\frac{\dot{\mathcal{V}}}{i\omega_0}) = \sqrt{\frac{C}{2}}|V|e^{-i\omega_0t}, \\
  a^* = \sqrt{\frac{C}{2}}(\mathcal{V} +i\sqrt{\frac{L}{C}}\mathcal{I}) = \sqrt{\frac{C}{2}}(\mathcal{V} +
\frac{\dot{\mathcal{V}}}{i\omega_0})=\sqrt{\frac{C}{2}}|V|e^{i\omega_0t}, 
  \end{array}
  \label{CMT4}
\end{equation}
The voltage $\mathcal{V}(t)$ and its derivative $\mathcal{\dot{V}}(t)$ can then be expressed in terms of the 
complex mode amplitudes as: 
\begin{equation}
\centering
 \begin{array}{ccc}
\mathcal{V} = \frac{1}{\sqrt{2C}}(a+a^*)\\
\mathcal{\dot{V}} = \frac{-i\omega_0}{\sqrt{2C}}(a-a^*).
\end{array}
\label{CMT5}
\end{equation}
from where the expression for the current follows immediately using the second equation of Eq. (\ref{CMT2}).

We are now moving forward with the description of the $\mathcal{PT}$ symmetric system of coupled $RLC$ 
circuits (Fig.1a of the main text) using the complex amplitudes representation. First, we write Kirchhoff's laws 
for the voltages at the nodes of the $RLC$ resonators. They take the form
\begin{equation}
\centering
 \Bigg\{ \begin{array}{ccc}
(1+\kappa)\mathcal{\ddot{V}}_1-\gamma_1\omega_0\mathcal{\dot{V}}_1+\omega_0^2\mathcal{V}_1-\kappa
\mathcal{\ddot{V}}_2=0\\
(1+\kappa)\mathcal{\ddot{V}}_2+\gamma_2\omega_0\mathcal{\dot{V}}_2+\omega_0^2\mathcal{V}_2-\kappa
\mathcal{\ddot{V}}_1=0, 
  \end{array}
  \label{CMT6}
\end{equation}
where $\gamma_{1,2}=\frac{1}{R_{1,2}}\sqrt{\frac{L}{C}}$ is the gain/loss of the first/second $LC$ resonator due to 
the negative/positive resistance $\mp R_{1,2}$ and $\kappa = \frac{C_c}{C}$ is the ratio of the coupling capacitor $C_c$ 
to the capacitance $C$, which represents the coupling strength between the two resonators. Equation (\ref{CMT6}) 
can be rewritten in terms of the complex amplitudes $a_1,a_2$ and their conjugates $a_1^*,a_2^*$ as \tp{\cite{HAUS,FKLK21}}
\begin{equation}
\centering
 \Bigg\{ \begin{array}{ccc}
\frac{d}{dt}a_1 \approx -i\omega_0a_1+\frac{1}{2}\omega_0\gamma_1 (a_1-a_1^*)+\frac{1}{2}i\omega_0\kappa(a_1-a_1^*-a_2+a_2^*)\\
\frac{d}{dt}a_2 \approx -i\omega_0a_2-\frac{1}{2}\omega_0\gamma_2 (a_2-a_2^*)+\frac{1}{2}i\omega_0\kappa(a_2-a_2^*-a_1+a_1^*), 
  \end{array}
  \label{CMT7}
\end{equation}
where it was assumed that the coupling between the resonators is weak i.e. $\kappa\ll1$. We can simplify further the above 
equation by invoking a rotating-wave approximation. This allows us to decouple $a$ and $a^*$ and re-write Eq. (\ref{CMT7}) 
in a CMT form: 
\begin{equation}
\centering
 \frac{d}{dt}\mathbf{a} \approx -i\mathbf{H}_0\mathbf{a}, \,\,\,\mathbf{H}_0 =\omega_0\mathbf{\widetilde{H}}_0= \omega_0\ \left( \begin{array}{ccc}
1-\frac{\kappa}{2}+i\frac{\gamma_1}{2} & \frac{\kappa}{2}  \\
\frac{\kappa}{2} & 1-\frac{\kappa}{2}-i\frac{\gamma_2}{2}  \end{array} \right),\,\,\,\mathbf{a}=(a_1,a_2)^T
  \label{CMT8}
\end{equation}
where $\mathbf{\widetilde{H}}_0$ is the (dimensionless) Hamiltonian of the two-mode system (below a tilde will indicate 
dimensionless operators). 

Next, we incorporate in our modeling the (weak) coupling of the $LC$ resonators to the transmission lines occurring 
via the capacitance $C_e=\kappa_eC$, $\kappa_e\ll1$. At the transmission lines the voltage $\mathcal{V}_{\rm TL}$ 
and the current $\mathcal{I}_{\rm TL}$ are represented as a superposition of a forward (+) and a backward (-) propagating 
wave i.e. $\mathcal{V}_{\rm TL} =\mathcal{V}_{\rm TL}^++\mathcal{V}_{\rm TL}^-$, $\mathcal{I}_{\rm TL} =
\mathcal{I}_{\rm TL}^++\mathcal{I}_{\rm TL}^- = \frac{1}{Z_0}(\mathcal{V}_{\rm TL}^+-\mathcal{V}_{\rm TL}^- )$, where 
$Z_0$ is a characteristic impedance of the transmission line. At the same time we express the voltages in terms of 
complex wave amplitudes as: $\mathcal{V}_{\rm TL}^{\pm}=\pm\sqrt{\frac{Z_0}{2}}(S^{\pm}+S^{\pm*})$, where $S^{\pm} 
= |S^{\pm}| e^{-i\omega_0t}$. The Kirchhoff's equations that describe the voltage and current at the junction between 
the transmission line and a resonator with loss given by $\gamma_0$ is:
\begin{equation}
\frac{1}{C}\frac{d}{dt}\mathcal{I}_{\rm TL} = \mathcal{\ddot{V}}+\gamma_0\omega_0 \mathcal{\dot{V}}+\omega_0^2 \mathcal{V};\quad
\mathcal{I}_{\rm TL}=C_e\frac{d}{dt}\left(\mathcal{V}_{\rm TL}-\mathcal{V}\right). 
  \label{CMT9}
\end{equation}
Furthermore, assuming weak coupling $\kappa_e\ll1$ and considering a rotating-wave approximation, we can re-write 
the above set of equations in the form:
\begin{equation}
\centering
 \Bigg\{ \begin{array}{ccc}
\frac{d}{dt}a \approx -i\omega_0 \left(1-\frac{1}{2}\gamma_e\right)a-\frac{1}{2}\omega_0\gamma_0a-\frac{1}{2}\omega_0\gamma_ea-i\sqrt{\omega_0\gamma_e}S^{+}\\
S^{-}\approx S^+-i\sqrt{\omega_0\gamma_e}a. 
  \end{array}
  \label{CMT10}
\end{equation}
where $\gamma_e=\frac{Z_0}{\sqrt{L/C}}\kappa_e^2$. 

Combining together Eqs. (\ref{CMT8},\ref{CMT10}) allows us to describe our RLC circuit in a temporal CMT form, where after redefinition of phase factors of the  propagated waves in the transmission lines as $\mathcal{S}^{\pm} \xrightarrow[]{} \pm i\mathcal{S}^{\pm}$. we get:
\begin{equation}
\centering
 \Bigg\{ \begin{array}{ccc}
\frac{d}{dt}\mathbf{a} = -i(\mathbf{H}_0-\frac{i}{2}\mathbf{W}^T\mathbf{W})\mathbf{a}+\mathbf{W}^T\mathbf{\mathcal{S}}^+\\
\mathbf{\mathcal{S}}^{-}=-\mathbf{\mathcal{S}}^{+}+\mathbf{Wa}.\,\,\, 
  \end{array},\,\,\,\,\,\,\mathbf{W}\equiv\sqrt{\omega_0}\mathbf{\widetilde{W}} =  \left( \begin{array}{ccc}
\sqrt{\omega_0\gamma_e} & 0  \\
0 &\sqrt{\omega_0\gamma_e}  \end{array} \right),\,\,\mathcal{S}^{\pm}=(S^{\pm1}_1, S^{\pm}_2)^T
  \label{CMT11}
\end{equation}

Using Eq. (\ref{CMT8}) and Eq. (\ref{CMT11}) we are now able to map the electronic circuit used in 
our experiment to the CMT formalism which allows to generalize the analysis of our platform the broad range of frameworks. 
In the sections below we proceed with the general CMT analysis. For convenience we have redefined the various parameters 
associated with the LRC circuit as: $\omega_0(1-\frac{\kappa}{2}-\frac{\kappa_e}{2})\rightarrow \omega_0$, $\kappa \rightarrow 
2\kappa$, $\gamma_{1,2} \rightarrow 2\gamma_{1,2}$ and $\gamma_e \rightarrow 2\gamma_e$. Moreover, since 
$\gamma_e$ and $\kappa_e$ are directly related (see above), we will be referring to the isolated ${\cal PT}$-dimer (zero 
coupling with the transmission lines i.e. $\kappa_e=0$) with the equivalent condition $\gamma_e=0$.

\section{Analysis of the CMT Model and Scattering Approach}

We first analyze the eigen-spectrum of the Hamiltonian $\mathbf{\widetilde{H}}_0$ (see previous section) which 
describes the isolated system (i.e. $\gamma_e=0$) :
\begin{equation}
\mathbf{\widetilde{H}}_0 \equiv\frac{\mathbf{H}_0}{\omega_0}\ = \ \left( \begin{array}{ccc}
1+i\gamma_1 & \kappa  \\
\kappa & 1-i\gamma_2  \end{array} \right);
\label{S1}
\end{equation} 
where $\gamma_{1,2}$ are the dimensionless linewidths of the two resonant modes due to the presence of the gain/loss 
elements at the first and the second LC resonator respectively. In case 
of ${\cal PT} $ symmetric configurations the gain/loss elements are perfectly balanced and therefore $\gamma_1 = 
\gamma_2=\gamma_0$.  

Direct diagonalization of Eq.~(\ref{S1}) allows us to evaluate the (dimensionless) eigenfrequencies and the eigenvectors 
of the coupled mode system:
\begin{equation}
\widetilde{\omega}_{\pm}^{(0)} = \frac{\omega_{\pm}^{(0)}}{\omega_0}=1 \pm \sqrt{\kappa^2-\gamma^2_0};\quad \quad
{\bf\upsilon}_{\pm}^{(0)}=\left(\frac{i\gamma_0 \pm  \sqrt{\kappa^2-\gamma^2_0}}{\kappa},1\right)^T
\label{S2}
\end{equation} 
From Eq. (\ref{S2}) we conclude that the two modes of the ${\cal PT}$ circuit demonstrate an exceptional point (EP) degeneracy 
when the coupling constant $\kappa$ becomes equal to the gain/loss strength i.e. $\kappa_{EP}=\gamma_0$. For this 
coupling strength, the value of the degenerate frequency becomes $\widetilde{\omega}_{\pm}^{(0)}=\widetilde{\omega}_{EP}^{(0)}
=1$. 

Next, we assume that the couple resonators are initially positioned at the EP configuration while a small perturbation 
$\epsilon$, associated with a small variation at the coupling capacitor due to an applied acceleration, is imposed to 
the system. Such perturbations will affect the coupling rate as $\kappa=\kappa_{EP}\rightarrow \kappa_{EP}+\epsilon$. 
At the same time they will induce a shift at the natural frequencies of the resonators $\omega_0\rightarrow \omega_0
(1-\epsilon)$ (see Eq. (\ref{CMT8})). Finally, we take into account the linewidth broadening of the natural frequencies 
of the individual resonators due to their coupling to the transmission lines (see Eq. (\ref{CMT11})). Incorporating these 
perturbation effects in Eq. (\ref{S1}), we come up with the following (dimensionless) effective Hamiltonian (see Eq. 
(\ref{CMT11})) 
\begin{equation}
\begin{array}{ccc}
\widetilde{\mathbf{H}}_{\rm eff}=\mathbf{\widetilde{H}}_{0}-\frac{i}{2}\widetilde{\mathbf{W}}^{T}\widetilde{\mathbf{W}} =
\left( \begin{array}{ccc}
1-\epsilon-i\gamma_e+i \gamma_0 & \gamma_0+\epsilon  \\
\gamma_0+\epsilon & 1-\epsilon-i\gamma_e-i\gamma_0  \end{array} \right);\end{array}
\label{S03}
\end{equation} 
where the (dimensionless) linewidth broadening term $\gamma_e$ is introduced via the matrix $\widetilde{\mathbf{W}}$, see Eq. (\ref{CMT11}).
A direct diagonalization of $\widetilde{\mathbf{H}}_{\rm eff}$ gives us the eigenmodes of the open system
\begin{equation}
\widetilde{\omega}_{\pm}^{\rm eff} (\epsilon; \gamma_e)= 1-\epsilon-i\gamma_e \pm \sqrt{2\gamma_0\epsilon+\epsilon^2};
\quad\mathbf{\upsilon}_{\pm}^{\rm eff}=\left(\frac{i\gamma_0 \pm  \sqrt{2\gamma_0\epsilon+\epsilon^2}}{\gamma_0+\epsilon},
1\right)^T
\label{S5}
\end{equation} 
which, similar to the isolated system (see Eq. (\ref{S2})), they also have an EP degeneracy at $\epsilon=\epsilon_{EP}=0$. 
Notice that for $\gamma_e=0$, the eigenmodes of Eq. (\ref{S5}) coincide with the eigenmodes $\widetilde{\omega}_{\pm}^{(0)}
(\epsilon)$ of the Hamiltonian $\mathbf{\widetilde{H}}_0(\epsilon)=\mathbf{\widetilde{H}}_{\rm eff}(\epsilon;\gamma_e=0)$ 
describing the isolated ${\cal PT}$-symmetric dimer in the case that a perturbation $\epsilon$ has modified the coupling 
between the two resonators and their natural frequencies as discussed above.

The scattering matrix that describes the transport properties of our system is evaluated from Eq. (\ref{CMT11}) and takes 
the form
\begin{equation}
\mathbf{S}(\widetilde{\omega})=-\mathbf{I}-i\widetilde{\mathbf{W}}\mathbf{G}(\widetilde{\omega})
\widetilde{\mathbf{W}}^T;\quad \mathbf{G}(\widetilde{\omega})=(\widetilde{\mathbf{H}}_{\rm eff}
-\widetilde{\omega} \mathbf{I})^{-1}.
\label{S6}
\end{equation} 
where $\mathbf{I}$ is an identity matrix and $G(\widetilde{\omega})$ is the Green's function associated with the 
effective Hamiltonian $\widetilde{\mathbf{H}}_{\rm eff}$ of Eq. (\ref{S03}). A direct substitution of Eq. (\ref{S03}) into Eq. (\ref{S6}) 
allows us to evaluate the individual Green's function matrix elements:
\begin{equation}
\begin{array}{cc}
G_{11}(\widetilde{\omega}) =\frac{1-\epsilon-i\gamma_e-i\gamma_0-\widetilde{\omega}}{(-1-\widetilde{\omega}-i\gamma_e)^2+
2(1-\widetilde{\omega}-i\gamma_e-\gamma_0)\epsilon},  \\
G_{12}(\widetilde{\omega})=G_{21}(\widetilde{\omega}) =\frac{\gamma_0+\epsilon}{-(-1-\widetilde{\omega}-i\gamma_e)^2+
2(1-\widetilde{\omega}-i\gamma_e-\gamma_0)\epsilon}, \\
G_{22}(\widetilde{\omega}) =\frac{1-\epsilon-i\gamma_e+i\gamma_0-\widetilde{\omega}}{(-1-\widetilde{\omega}-i\gamma_e)^2
+2(1-\widetilde{\omega}-i\gamma_e-\gamma_0)\epsilon}. 
\end{array}
\label{S8}
\end{equation}
and via Eq. (\ref{S6}) the transmittance. The latter takes the form   
\begin{equation}
\mathcal{T}(\widetilde{\omega})\equiv|S_{21}(\widetilde{\omega})|^2 = \frac{4\gamma^2_e(\gamma_o+\epsilon)^2}
{(-1+\widetilde{\omega}-\epsilon)^4+2(\gamma^2_e-2\gamma_0\epsilon-\epsilon^2)(-1+\widetilde{\omega}+\epsilon)^2
+(\gamma^2_e+2\gamma_0\epsilon+\epsilon^2)^2}.
\label{S9}
\end{equation} 
which allow us to extract analytically the trajectories of the frequencies associated with the transmission peaks. The 
later are the physical observables that have been used in our sensing study (see main text) and they take the form:
\begin{equation}
\widetilde{\omega}_{\pm}=\left\{
\begin{array}{cc}
1-\epsilon;& \epsilon\leq\epsilon_{\rm TPD}  \\
1-\epsilon\pm\sqrt{2\gamma_0\epsilon+\epsilon^2-\gamma_e^2};& \epsilon\geq\epsilon_{\rm TPD}
\end{array}
\right.
\label{S10}
\end{equation}
where $\epsilon_{\rm TPD} = -\gamma_0+\sqrt{\gamma_0^2+\gamma_e^2}(\neq
\epsilon_{\rm EP}=0)$ is their coalescence point. At this perturbation strength we have that $\widetilde{\omega}_{\pm}
=\widetilde{\omega}_{\rm TPD}=1-\epsilon_{\rm TPD}(\neq\widetilde{\omega}_{\rm EP}
=1)$. From the above equation it is straightforward to show that the transmission peak frequencies around the transmission 
peak degeneracy (TPD) point, follow a sublinear Puiseux expansion with a fractional $1/2$ power. This is a consequence 
of the EP degeneracy of the eigenfrequencies of the underlying isolated system. It is crucial to point out that while 
$\widetilde{\omega}_{\pm}$ follow closely the trajectories of the corresponding eigenfrequencies $\widetilde{\omega}_{\pm}^{(0)}$, 
their degeneracy occurs at a 
different parameter value i.e. $\epsilon_{\rm TPD}>\epsilon_{\rm EP}$. As a result, at the TPD the bi-orthogonal basis 
of the effective Hamiltonian Eq. (\ref{S03}) does not collapse and, therefore, the associated Petermann factor does not 
diverge. This divergence of the Petermann factor was considered the source of the sensitivity limitations imposed to 
the Brillouin ring laser gyroscope in Ref.\cite{YWWGV20}. Furthermore, the separation of $\epsilon_{\rm TPD}$ and $\epsilon_{\rm 
EP}$ guarantees the analyticity of the various physical observables in the proximity of TPD.

Furthermore, from Eqs. (\ref{S9},\ref{S10}) we can calculate the scaling of the transmittance peak $\mathcal{T}_{peak}$ as 
follows:
\begin{equation}
\mathcal{T}_{peak}=\left\{
\begin{array}{cc}
\frac{4\gamma_e^2(\gamma_0+\epsilon)^2}{\left(\gamma_e^2+2\gamma_0\epsilon+\epsilon^2\right)^2}; & \epsilon\leq\epsilon_{\rm TPD} \\
\frac{(\gamma_0+\epsilon)^2}{2\gamma_0\epsilon+\epsilon^2};& \epsilon\geq\epsilon_{\rm TPD}
\end{array}
\right.
\label{S101}
\end{equation}
This implies that at the point of transmittance peak coalescence $\epsilon=\epsilon_{TPD} $, the transmittance peak approaches 
the value $\mathcal{T}_{peak}(\epsilon_{TPD}) = 1+\left(\frac{\gamma_0}{\gamma_e}\right)^2$.

Finally, from Eq. (\ref{S6},\ref{S8}) we have identified the lasing conditions of the open system, as the frequency $\widetilde{
\omega}_L$ 
and perturbation $\epsilon_L$ values for which the scattering matrix diverges. The corresponding values for lasing action are
$\epsilon_L=-\gamma_0+ \sqrt{\gamma_0^2-\gamma_e^2}$, and $\omega_L=1-\epsilon_L$. We have, therefore, that: 
$\epsilon_L<\epsilon_{EP}=0<\epsilon_{\rm TPD}$.  

A panorama of the transmittance spectrum versus the perturbation strength $\epsilon$ is shown as density plots for various 
representative values of $\gamma_e>\gamma_0$ in Figs. \ref{figs2}a-d. At the same figures we also report the trajectory  
of the transmission peaks (green dashed lines) Eq. (\ref{S10}), together with the eigenmodes ${\cal R}e\left(\widetilde{
\omega}_{\pm}^{\rm eff}(\epsilon;\gamma_e)\right)$ of the Hamiltonian $\widetilde{\textbf{H}}_{\rm eff}$, see Eq. (\ref{S5}).
We point out that these modes are the same with the eigenmodes of $\widetilde{\textbf{H}}_{0}(\epsilon)$, and therefore 
we will not distinguish them in this work.

\section{Quantifications of Sensitivity Enhancement and Non-Orthogonality of modes}

In order to quantify the non-orthogonal nature of the eigenmodes of our system, and specifically its proximity to an EP, we 
are introducing the so-called Petermann factor ($PF$) \cite{YWWGV20}. 
Specifically:  
\begin{equation}
PF(\epsilon) =\frac{1}{2}\left( 1+\frac{Tr(\mathbf{H}_{tr}^+\mathbf{H}_{tr})}{\left|Tr(\mathbf{H}_{tr}^2)\right|}\right) = 
\frac{(\gamma_0+\epsilon)^2}{2\gamma_0\epsilon+\epsilon^2}, 
\label{S12}
\end{equation}
where $\mathbf{H}_{tr}=\mathbf{H}_{\rm eff}-\frac{1}{2}Tr(\mathbf{H}_{\rm eff})$ is a traceless part of the Hamiltonian 
$\mathbf{H}_{\rm eff}$. Notice that the $PF(\epsilon)$ evaluated in Eq. (\ref{S12}) is the same one associated with the 
$\mathbf{H}_{\rm eff}(\epsilon;\gamma_e=0)\equiv \mathbf{H}_0(\epsilon)$ and therefore we do not distinguish these 
two below. We will show that the Petermann factor $PF$ is related with the so-called noise enhancement factor (NEF) 
of our ${\cal PT}$-symmetric circuit.

Furthermore, in order to quantify the sensitivity enhancement associated with the square-root degeneracy of the eigenfrequencies 
of the system ($SEF_{\rm mode}$), we have intrioduced the sensitivity enhancement factor
\begin{equation}
SEF_{mode}(\epsilon) \equiv\left|\frac{\partial(\Delta\widetilde{\omega}^{\rm eff})}{2\partial \epsilon}\right|^2 
= \frac{(\gamma_0+\epsilon)^2}{2\gamma_0\epsilon+\epsilon^2},
\label{S11}
\end{equation}
where $\Delta\widetilde{\omega}^{\rm eff} = (\widetilde{\omega}_{+}^{\rm eff}-\widetilde{\omega}_{-}^{\rm eff})$. The final 
equality in Eq. (\ref{S11}) requires 
to substitute the expressions for the resonant frequencies of the $\widetilde{\mathbf{H}}_{\rm eff}$ from Eq. (\ref{S5}). It turns out 
that both $PF(\epsilon)$ and $SEF_{\rm mode}(\epsilon)$ are given by the same expressions.

Since our sensing protocol utilizes the transmission peak frequencies $\widetilde{\omega}_{\pm}$ for measuring the differential 
acceleration, it is natural to introduce an associated sensitivity enhancement factor which is given by:
\begin{equation}
SEF(\epsilon) =\left|\frac{\partial(\Delta\widetilde{\omega})}{2\partial \epsilon}\right|^2 = \frac{(\gamma_0+\epsilon)^2}{2\gamma_0
\epsilon+\epsilon^2-\gamma_e^2}.
\label{S13}
\end{equation}
where $\Delta\widetilde{\omega}=\widetilde{\omega}_{+}-\widetilde{\omega}_{-}$.
From the above expression we see that the $SEF$ diverges at $\epsilon_{\rm TPD} = -\gamma_0+\sqrt{\gamma_0^2+
\gamma_e^2}$ where the transmission peaks coalesce. 

These three quantities are reported in Fig. \ref{figs2} for three representative values of linewidth broadening $\gamma_e$ due to 
the coupling of the ${\cal PT}$-symmetric circuit to the transmission lines.

\section{ Thermal noise associated with the transmission line and the gain/loss elements of the $\mathcal {PT}$-symmetric circuit}

The transport properties our ${\cal PT}$-symmetric circuit (see Fig. \ref{figs1}) in the presence of noise sources can be analyzed 
using a temporal CMT \cite{HAUS,FAN}:
\begin{equation}
\centering
\Bigg\{ \begin{array}{ccc}
  \frac{d\mathbf{a}}{d\widetilde{t}}=-i\widetilde{\mathbf{H}}_{\rm eff}\mathbf{a}+\mathbf{K}^T\mathbf{n}^{in} \\
  \mathbf{n}^{out}=\mathbf{Ka+Cn}^{in},
  \end{array}
  \label{S14}
\end{equation}
where $\widetilde{t}=\omega_0\cdot t$ is the dimensionless time, $\widetilde{\mathbf{H}}_{\rm eff}$ is the effective Hamiltonian given by Eq. (\ref{S03}), 
$\mathbf{K}$ is a matrix modeling the coupling between the system and the various noise sources involved in the problem and 
$\mathbf{C}$ is a matrix that describes a direct scattering process and has to satisfy the relation $\mathbf{CK}^*=-\mathbf{K}$ \cite{FAN}. 
In the scenario that we consider here, the coupling matrices $\mathbf{K}$ and $\mathbf{C}$ have the form: 
\begin{equation}
\centering
\mathbf{K}=\left( \begin{array}{ccc}
  \sqrt{2\gamma_0} & 0 \\
 0 & \sqrt{2\gamma_0} \\
 \sqrt{2\gamma_e} & 0 \\
0 & \sqrt{2\gamma_e}  
  \end{array}
\right), 
  \mathbf{C}=\left( \begin{array}{cccc}
  -1 & 0 & 0 & 0 \\
 0 & -1 & 0 & 0 \\
 0 & 0 & -1 & 0 \\
0 & 0 & 0 & -1  
  \end{array}
\right) .
  \label{S15}
\end{equation}
Finally, in Eq. (\ref{S14}) the state vector $\mathbf{a}$ describes the excitation inside the scattering domain, and 
$\mathbf{n}^{in} (\mathbf{n}^{out})$ are state vectors describing incoming (outgoing) excitations from the noise sources 
to the system (from the system to the noise sources, including the transmission lines). 

We assume that there are four noise sources whose inputs are described by the vector $\mathbf{n}^{in}$ $ = (n_{1}^{in}, 
n_{2}^{in}, n_{{\rm TL}_1}^{in}, n_{{\rm TL}_2}^{in})^T$. The vector components $n_{1/2}^{in}$ describe noise signals from 
the gain/loss sources which are connected to the left/right resonator of the system respectively, while the vector components 
$n_{{\rm TL}_1/{\rm TL}_2}^{in}$ describe incident noise signals associated with the left/right transmission lines. The properties 
of the noise sources are described by their correlations 
\begin{eqnarray}
\langle n_{{\rm TL}_l}^{*in}(\widetilde{\omega}) n_{{\rm TL}_m}^{in}(\widetilde{\omega}')\rangle &=& 4k_bT_{{\rm TL}_l}
Z\Delta_f\delta_{lm}
\delta(\widetilde{\omega}-\widetilde{\omega}')\\
\langle n_l^{*in}(\widetilde{\omega}) n_m^{in}(\widetilde{\omega}')\rangle &=& 4k_bT_lR\Delta_f\delta_{lm}\delta(
\widetilde{\omega} -\widetilde{\omega}')\nonumber
\end{eqnarray}
where $k_b$ is the Boltzmann cnstant, $T_{{\rm TL}_l}$ is the temperature of the noise reservoir associated with the 
transmission line $l$, $T_{l}$ is the temperature of the noise source associated with the gain/loss elements at the resonator 
$l$, $Z$ is the impedance of the transmission 
lines, $R$ is the resistance responsible for the amplification/attenuation mechanisms associated with the two modes of the 
system, and $\Delta_f$ - is the frequency bandwidth over which the signal at each frequency is measured (i.e. $\Delta_f=
1/\tau$ is inversely proportional to the sampling time $\tau$ over which the signal is measured and 
averaged). Finally, $\delta_{ij}$ and $\delta(x)$ are the Kronecker delta and the Dirac delta functions respectively.

The output noise signal $\textbf{n}^{out}$ is evaluated from Eq. (\ref{S14}) and takes the form 
\begin{equation}
\mathbf{n}^{out} = \left( \begin{array}{cccc}
n_1^{out}\\
n_2^{out}\\
n_{{\rm TL}_1}^{out}\\
n_{{\rm TL}_2}^{out}
\end{array}\right) = -i\mathbf{KGK}^T +\mathbf{Cn}^{in},
\label{S16}
\end{equation}
where $\mathbf{G}$ is the Green's function given by Eq. (\ref{S6}). The vector component $n_l^{out}$ describes the noise 
output detected at the $l$-resonator of the system and $n_{{\rm TL}_l}^{out}$ describes the noise output emitted at the $l-$th 
transmission line (see Fig. \ref{figs1}). Below we analyze the effects of the output noise signal at the right transmission 
line $n_{{\rm TL}_2}^{out}$, on the transmission measurements (we assume that the incident signal is from the left transmission 
line). The right noise output is given by
\begin{equation}
n_{{\rm TL}_2}^{out} = -n_{{\rm TL}_2}^{in}-i\sqrt{2\gamma_e}\left( \sqrt{2\gamma_e}G_{21}n_{{\rm TL}_1}^{in}  +
\sqrt{2\gamma_e}G_{22}n_{{\rm TL}_2}^{in} + \sqrt{2\gamma_0}G_{21}n_{1}^{in} + \sqrt{2\gamma_0}G_{22}n_{2}^{in} \right),
\label{S17}
\end{equation}
where $G_{lm}$ are the Green's function elements given by Eq. (\ref{S8}). Armed with this knowledge we are now 
able to evaluate the noise power associated with measurements at the right transmission line:
\begin{equation}
\mathcal{S}_{{\rm TL}_2}^{C}(\widetilde{\omega})  =  \langle \left|n_{{\rm TL}_2}^{out}(\widetilde{\omega})\right|^2\rangle = 
\mathcal{S}_{{\rm TL}_2}^{{\cal PT}}(\widetilde{\omega}) + \mathcal{S}_{{\rm TL}_2}^{TL}(\widetilde{\omega})
\label{S17b}
\end{equation}
where the super-index $C$ indicates the cumulative circuit noise from the gain/loss elements of the circuit (denoted 
with the super-index ${\cal PT}$) and the ambient noise originating from the transmission lines (denoted with the 
super-index $TL$). The contribution from the internal (external) noise sources $n_{1,2}^{in} 
(n_{{\rm TL}_{1,2}}^{in})$ are denoted as $\mathcal{S}_{{\rm TL}_2}^{{\cal PT}} (\mathcal{S}_{{\rm TL}_2}^{TL})$ and are given by: 
\begin{equation}
\begin{array}{cccc}
\mathcal{S}_{{\rm TL}_2}^{{\cal PT}} = 4k_BTR\Delta_f \left( 4\gamma_e\gamma_0\Big(|G_{22}|^2+|G_{21}|^2\right) \Big)\\
\mathcal{S}_{{\rm TL}_2}^{TL} = 4k_BTZ\Delta_f \Big(1-4\gamma_e\mathcal{I}m(G_{22})+4\gamma_e^2\left(|G_{22}|^2
+|G_{21}|^2\right)\Big).
\end{array}
\label{S18}
\end{equation}
Since our sensing scheme relies on measuring the frequency splitting between the transmission peaks $\Delta\widetilde{\omega}$, 
it is important the analyze the noise power at frequencies $\widetilde{\omega}_{\pm}$, for $\epsilon\geq \epsilon_{\rm TPD}$ 
(see Eq. (\ref{S10})), where its contribution will be detrimental. In this respect, we evaluate at $\widetilde{\omega}_{\pm}$ the 
Green's function elements from Eq. (\ref{S8}) and using Eq. (\ref{S18}), we get that:  
\begin{equation}
\mathcal{S}_{{\rm TL}_2}^{{\cal PT}}(\epsilon) = 4k_BTR\Delta_f \left(\frac{2\gamma_0}{\gamma_e}\frac{(\gamma_0+\epsilon)^2
-\gamma_0\gamma_e}{2\gamma_0\epsilon+\epsilon^2} \right);\quad
\mathcal{S}_{{\rm TL}_2}^{TL}(\epsilon) = 4k_BTZ\Delta_f \left(\frac{(\gamma_0+\epsilon)^2+\gamma_0^2}{2\gamma_0
\epsilon+\epsilon^2}\right)
\label{S19}
\end{equation}
which allow us to conclude that both $\mathcal{S}_{{\rm TL}_2}^{{\cal PT}}$ and $\mathcal{S}_{{\rm TL}_2}^{TL}$ saturate 
to a final value at perturbations $\epsilon=\epsilon_{\rm TPD}$ where the transmission peaks coalesce and the SEF diverges (see 
Eq. (\ref{S13})). 

Next, we evaluate the noise enhancement factors $NEF_{{\rm TL}_2}^{{\cal PT}}(\epsilon)$ and $NEF_{{\rm TL}_2}^{TL}(\epsilon)$, 
defined as the ratio of the noise power at a perturbation $\epsilon$ to the noise power evaluated at the $\epsilon\rightarrow\infty$ limit. 
In the domain $\epsilon\geq \epsilon_{\rm TPD}$ we get
\begin{equation}
NEF_{{\rm TL}_2}^{{\cal PT}}(\epsilon) \equiv \frac{\mathcal{S}_{{\rm TL}_2}^{{\cal PT}}(\epsilon)}{\mathcal{S}_{{\rm TL}_2}^{{\cal PT}}(\epsilon\rightarrow\infty)}
=\frac{(\gamma_0+\epsilon)^2-\gamma_e\gamma_0}{2\gamma_0\epsilon+\epsilon^2};\quad
NEF_{{\rm TL}_2}^{TL}(\epsilon) \equiv \frac{\mathcal{S}_{{\rm TL}_2}^{TL}(\epsilon)}{\mathcal{S}_{{\rm TL}_2}^{TL}(\epsilon\rightarrow\infty)}
=\frac{(\gamma_0+\epsilon)^2+\gamma_0^2}{2\gamma_0\epsilon+\epsilon^2}
\label{S21}
\end{equation}
where the asymptotic noise power values $\mathcal{S}_{{\rm TL}_2}^{{\cal PT}}(\epsilon\rightarrow\infty) = \frac{2\gamma_0}{\gamma_e} 
\cdot 4k_BTR\Delta_f$ and $\mathcal{S}_{{\rm TL}_2}^{TL}(\epsilon\rightarrow \infty) = 4k_BTZ\Delta_f$ have been evaluated 
directly from Eq. (\ref{S19}).

\begin{figure}
\centering
\includegraphics[width=1\columnwidth]{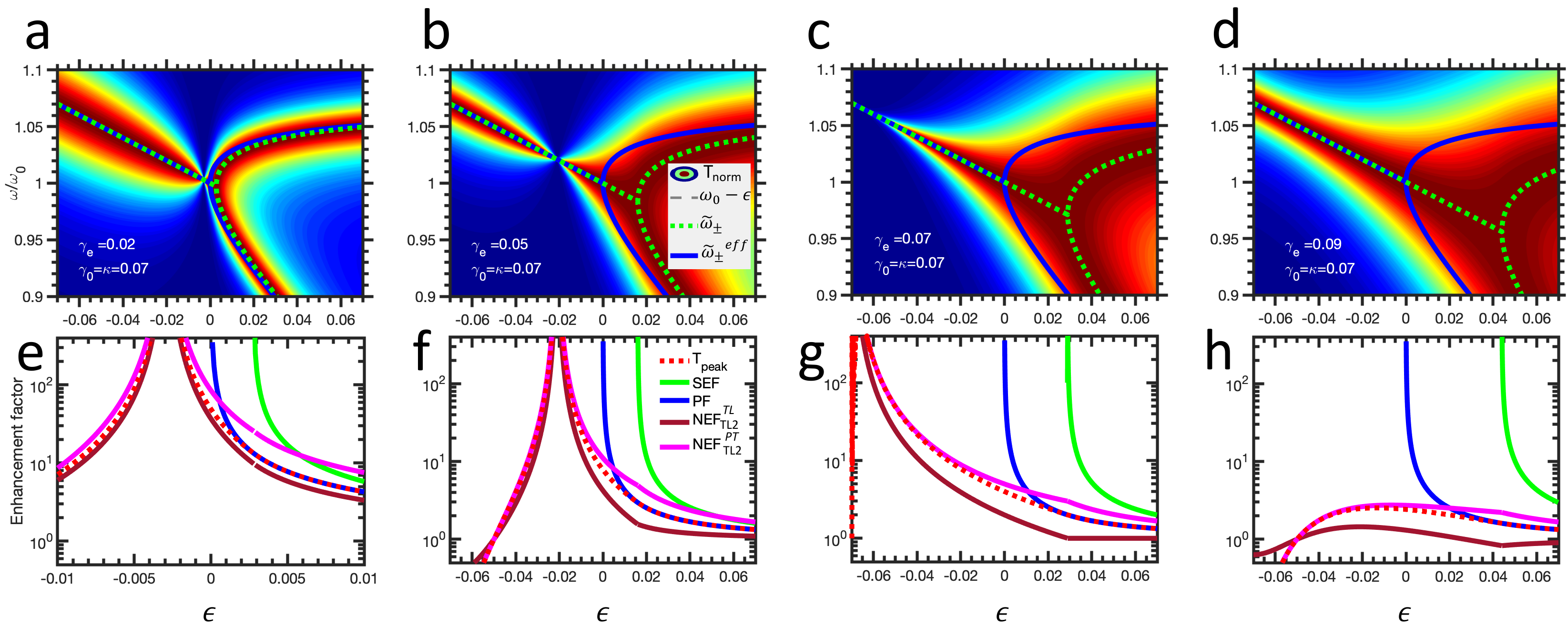}
\caption{\label{figs2} \linespread{0.5}\selectfont{} 
{\bf Spectral response of a $\mathcal{PT}$ symmetric dimer operating below the lasing threshold. } {\bf a-d}. 
Normalized transmission spectra as a function of the coupling variation $\epsilon$ for various values of coupling strength between 
the system and transmission lines $\gamma_e$. The blue solid lines denote the eigen-frequencies of the system ${\cal R}e\left( 
\omega_{\pm}^{\rm eff} \right)$ (see Eq. (\ref{S5})) and the green dotted lines denote the frequencies of the transmission peaks 
$\omega_{\pm}$. {\bf e-h.} The corresponding peak transmittance ${\mathcal T}_{peak}$ (red dotted line), $SEF$ (blue solid line), 
$NEF_{{\rm TL}_2}^{\cal PT}$ (dark red solid line) and $NEF_{{\rm TL}_2}^{TL}$ (magenta solid line) as a function of the coupling variation $\epsilon$}
\end{figure}

For the specific experimental parameters used in our set-up ($R\approx 1200$ Ohm, $Z \approx 50$ Ohm, and 
$\gamma_0> \gamma_e$) we can make further progress in the evaluation of the total noise enhancement factor. 
Specifically, we have that $\mathcal{S}_{{\rm TL}_2}^{\cal PT} \gg \mathcal{S}_{{\rm TL}_2}^{TL}$ (see Eq. (\ref{S19})), indicating that 
the noise enhancement factor $NEF_{{\rm TL}_2}^{C}(\epsilon)\equiv\left(\mathcal{S}_{{\rm TL}_2}^{\cal PT}(\epsilon)+
\mathcal{S}_{{\rm TL}_2}^{TL}
(\epsilon)\right)/\left(\mathcal{S}_{{\rm TL}_2}^{\cal PT}(\epsilon\rightarrow\infty)+\mathcal{S}_{{\rm TL}_2}^{TL}(\epsilon\rightarrow\infty)
\right)$ can be approximated as 
\begin{equation}
\label{S22}
NEF_{{\rm TL}_2}^{C}(\epsilon)\approx NEF_{{\rm TL}_2}^{\cal PT} = \frac{2\gamma_e^2((\gamma_0-\gamma_e)^2+(\gamma_0+\epsilon)^2)}{(\gamma_e^2+2\gamma_0\epsilon+\epsilon^2)^2}=PF(\epsilon)-\frac{\gamma_e\gamma_0}{2\gamma_0\epsilon+
\epsilon^2};\quad \epsilon\geq \epsilon_{\rm TPD}
\end{equation}
where $PF(\epsilon)$ is the Petermann factor, see Eq. (\ref{S12}). In fact, Eq. (\ref{S22}) can be further approximated 
at the weak coupling limit to the transmission lines $\gamma_e\ll\gamma_0$. In this case, the noise enhancement 
factor approaches the Petermann factor $PF$ (see Fig. \ref{figs2}), i.e.
\begin{equation}
NEF_{{\rm TL}_2}^{C}(\epsilon\geq\epsilon_{\rm TPD}, \gamma_0\gg\gamma_e ) \approx  PF(\epsilon);\quad \epsilon\geq \epsilon_{\rm TPD}
\label{S26}
\end{equation}

For the completeness of the study, we also report the noise enhancement factor in the parameter range $\epsilon<
\epsilon_{\rm TPD}$. In this case we have that the noise power (see Eq. (\ref{S18})) becomes:
\begin{eqnarray}
\mathcal{S}_{{\rm TL}_2}^{\cal PT} &=& 4k_BTR\Delta_f \left(\frac{4\gamma_0\gamma_e((\gamma_0-\gamma_e)^2+(\gamma_0
+\epsilon)^2)}{(\gamma_e^2+2\gamma_0\epsilon+\epsilon^2)^2} \right)\\
\mathcal{S}_{{\rm TL}_2}^{TL} &=& 4k_BTZ\Delta_f \left(1+\frac{4\gamma_0\gamma_e(2\gamma_0-\gamma_e+\epsilon)
(\gamma_0+\epsilon)}{(\gamma_e^2+2\gamma_0\epsilon+\epsilon^2)^2}\right)\nonumber
\end{eqnarray}
and consequently the corresponding noise enhancement factors are
\begin{equation}
NEF_{{\rm TL}_2}^{\cal PT}(\epsilon) =  \frac{2\gamma_e^2((\gamma_0-\gamma_e)^2+(\gamma_0+\epsilon)^2)}{(\gamma_e^2+
2\gamma_0\epsilon+\epsilon^2)^2};\quad
NEF_{{\rm TL}_2}^{TL}(\epsilon)  =   1+\frac{4\gamma_0\gamma_e(2\gamma_0-\gamma_e+\epsilon)(\gamma_0+\epsilon)}
{(\gamma_e^2+2\gamma_0\epsilon+\epsilon^2)^2}
\label{S24}
\end{equation}
The dependence of $NEF_{{\rm TL}_2}^{\cal PT}$ and $NEF_{{\rm TL}_2}^{TL}$ from the perturbation $\epsilon$ is shown in 
Fig. \ref{figs2}e-h for three typical cases of $\gamma_e<\gamma_0$. In all cases, the noise enhancement factor 
does not offsets the sensitivity enhancement factor $SEF$ at the proximity of the $\epsilon_{\rm TPD}$ in agreement 
with the conclusions of our experimental analysis (see main text).

\section{Uncertainty in the measurement of transmission peaks splitting due to cumulative circuit noise}

To properly estimate the sensor performance, one needs to study the amplification of the transmitted signal, in conjunction with the 
noise generated from the various sources. First we realize that the uncertainty $\sigma_{\Delta \widetilde{\omega}}^{C}$ in the measurement of the 
transmission peaks splitting $\Delta\widetilde{\omega}$, due to the cumulative circuit noise generated by the gain/loss elements of the circuit 
and the ambient noise of the transmission lines, is proportional to the spectral width of the transmission peaks $\Gamma$. Of course, 
the proportionality factor between $\sigma_{\Delta \widetilde{\omega}}^{C}$ and $\Gamma$ depends on a number of other parameters associated 
with the individual frequency sweeps (e.g. frequency step, spectral bandwidth, number of points, individual frequency bandwidth, 
period of individual frequency sweep, delay between the spectral sweeps, etc) during which a single transmission spectrum measurement 
is collected. Since we are not able to estimate all these contributions we are proceeding by assuming that this proportionality factor 
is mainly dominated by the noise produced by the gain/loss elements of the circuit and the ambient noise of the transmission lines. A 
closer inspection of Eq. (\ref{S9}) reveals that the spectral width is
\begin{equation}
\begin{array}{cccc}
\Gamma(\epsilon\geq \epsilon_{TPD})  &= &\sqrt{-\gamma_{e}^2+2\gamma_0\epsilon+\epsilon^2+2\gamma_e\sqrt{2\gamma_0
\epsilon+\epsilon^2}}-\sqrt{-\gamma_{e}^2+2\gamma_0\epsilon+\epsilon^2}\\ 
\Gamma(\epsilon\leq \epsilon_{TPD})  &= &\sqrt{-\gamma_{e}^2+2\gamma_0\epsilon+\epsilon^2+\sqrt{2}\sqrt{\gamma_e^4+\epsilon^2
(2\gamma_0+\epsilon)^2}}.
\end{array}
\label{S261}
\end{equation}

The uncertainty in the measurement of the transmission peak splitting  $\Delta\widetilde{\omega}$, due to the generated cumulative circuit noise, 
is then evaluated using the results from Eq. (\ref{S17b}):
\begin{equation}
\left(\sigma^{C}_{\Delta \widetilde{\omega}}\right)^2 \propto\Gamma^2(\epsilon) \frac{\mathcal{S}_{{\rm TL}_2}^{\cal PT}+ 
\mathcal{S}_{{\rm TL}_2}^{TL}}
{|S_{21}(\widetilde{\omega}=\widetilde{\omega}_{\rm TPD})|^2|\mathbf{a}_{{\rm TL}_1}^{in}|^2}= \left(\sigma_{\Delta\widetilde{\omega}}^{\cal PT}
(\epsilon)\right)^2+\left(\sigma_{\Delta\widetilde{\omega}}^{TL}(\epsilon)\right)^2,
\label{S27}
\end{equation}
where we have introduced the partial uncertainties $\sigma_{\Delta\widetilde{\omega}}^{\cal PT}(\epsilon)$ and $\sigma_{\Delta
\widetilde{\omega}}^{TL}(\epsilon)$ associated with the gain/loss power noise from the circuit and with the ambient noise from the 
transmission lines. The expression in the denominator in Eq. (\ref{S27}) is the output field intensity $\mathbf{a}_{L2}^{out}=S_{21}
\mathbf{a}_{L1}^{in}$ at the right transmission line, where we have assumed an input signal from the 
left transmission line $\mathbf{a}_{L1}^{in}$ and $S_{21}$ is the element of the scattering matrix given by Eq. (\ref{S9}). 
Obviously the above expression Eq. (\ref{S27}) applies for the perturbation range $\epsilon\geq 
\epsilon_{\rm TPD}$ where the transmission spectrum demonstrates two peaks. In this domain the noise powers $\mathcal{S}_{L2}^{\cal PT}$ 
and $\mathcal{S}_{L2}^{TL}$ are given by Eq. (\ref{S19}). Finally, the value of the transmission coefficient at the position of the peaks in this 
parameter domain, is found from Eq. (\ref {S9}) to be:
\begin{equation}
|S_{21}(\widetilde{\omega}=\widetilde{\omega}_{\pm})|^2=\frac{(\gamma_0+\epsilon)^2}{2\gamma_0\epsilon+\epsilon^2};\quad 
\epsilon\geq \epsilon_{\rm TPD}
\label{S28}
\end{equation}
which indicates that the signal transmitted from the left to the right transmission line is enhanced  in a same manner as a Petermann 
Factor (PF) of the isolated system (see Eq. (\ref{S12}) and Fig. \ref{figs2}). Furthermore, substituting in Eq. (\ref{S27}) the expressions 
for $\mathcal{S}_{L_2}^{\cal PT}, \mathcal{S}_{L_2}^{TL}$ from Eq. (\ref{S19}) we get
\begin{equation}
\begin{array}{cccc}
 \left(\sigma_{\Delta\widetilde{\omega}}^{\cal PT}(\epsilon)\right)^2 \propto\Gamma^2(\epsilon) \frac{4k_BTR\Delta_f}{|\mathbf{a}_{L1}^{in}|^2}
\cdot \frac{2\gamma_0}{\gamma_e}\left(1-\frac{\gamma_0\gamma_e}{(\gamma_0+\epsilon)^2} \right)\\ 
\left(\sigma_{\Delta\widetilde{\omega}}^{TL}(\epsilon)\right)^2 
\propto \Gamma^2(\epsilon) \frac{4k_BTZ\Delta_f}{|\mathbf{a}_{L1}^{in}|^2}\cdot \left(1+\frac{\gamma_0^2}{(\gamma_0+\epsilon)^2} 
\right).
 \end{array}
\label{S30}
\end{equation}
One can also introduce the uncertainty enhancement factor which describes the degree of enhancement in the uncertainty measurements 
of the transmission peak frequency splittings with respect to system configurations away from the TPD. We have 
\begin{equation}
\begin{array}{cccc}
\left(\Sigma_{\Delta\widetilde{\omega}}^{\cal PT}(\epsilon)\right)^2= \left(\frac{\sigma_{\Delta\widetilde{\omega}}^{\cal PT}(\epsilon)}
{\sigma_{\Delta\widetilde{\omega}}^{\cal PT}
(\epsilon\rightarrow\infty)}
\right)^2\propto \left(\frac{\Gamma^2(\epsilon)}{\gamma^2_e}\right)\left(1-\frac{\gamma_0\gamma_e}{(\gamma_0+\epsilon)^2} \right)\\ 
\left(\Sigma_{\Delta\widetilde{\omega}}^{TL}(\epsilon)\right)^2=\left(\frac{\sigma_{\Delta\widetilde{\omega}}^{TL}(\epsilon)}{\sigma_{\Delta
\widetilde{\omega}}^{TL}(\epsilon\rightarrow\infty)}\right)^2 
\propto \left(\frac{\Gamma^2(\epsilon)}{\gamma^2_e}\right)\left(1+\frac{\gamma_0^2}{(\gamma_0+\epsilon)^2} \right).
 \end{array}
\label{S32}
\end{equation}
where the ``asymptotic'' uncertainties have been calculated from Eq. (\ref{S30}) to be $\left(\sigma_{\Delta\widetilde{\omega}}^{\cal PT}
(\epsilon\rightarrow\infty)\right)^2 \propto\gamma_e^2 \frac{4k_BTR\Delta_f}{|\mathbf{a}_{L_1}^{in}|^2}\cdot \frac{2\gamma_0}{\gamma_e}$ 
and $\left(\sigma_{\Delta\widetilde{\omega}}^{TL}(\epsilon\rightarrow\infty)\right)^2 \propto \gamma_e^2 \frac{4k_BTZ\Delta_f}{|\mathbf{a}_{L_1}^{in}|^2}$.

Since our experimental platform operates under weak coupling conditions $\gamma_e\leq\gamma_0\ll1$ (units of $\omega_0$), we deduce 
from the above equation that both enhancement factors $\Sigma_{\Delta\widetilde{\omega}}^{\cal PT}(\epsilon)$ and 
$\Sigma_{\Delta\widetilde{\omega}}^{TL}(\epsilon)$ 
experience small variations in the proximity of $\epsilon\approx \epsilon_{\rm TPD}$, see Fig. \ref{figs3}a. We conclude, therefore, that the noise 
enhancement of the proposed platform leaves unaffected the uncertainty of the measurement of the transmission peak frequency splitting in 
the vicinity of $\epsilon=\epsilon_{\rm TPD}$ and does not offset the sensitivity enhancement, which diverges at the same point, see Eq. (\ref{S13}). 
This can be contrasted with the results of Ref. \cite{LLSYV19,YWWGV20} where they have found that the improved responsivity of a laser 
gyroscope operating near an EP is precisely compensated by increased laser noise.

\section {Uncertainty in the measurement of transmission peak splitting due to fluctuations of the coupling strength between the two 
modes}

Apart from the noise originated from the transmission lines and the gain/loss elements of the electronic circuit, our sensing 
platform suffers also from noise associated with fluctuations at the variation $\epsilon$ of the coupling constant $\kappa$ 
between the two resonators. These fluctuations are characterized by an uncertainty $\sigma_{\epsilon}^{\kappa}$ in the 
coupling variation $\epsilon$. Their physical origin ranges from the thermal motion of the test-mass (which is used to perturb 
the coupling strength -- coupling capacitance in our circuit) when external acceleration is applied, to fluctuations of the coupling 
capacitance due to thermal expansion of the glass plates used in our platform, or to variations in their dielectric 
properties. This uncertainty will be enhanced by the SEF given by Eq. (\ref{S13}), thus resulting in a cumulative coupling uncertainty
$\left(\sigma_{\Delta\widetilde{\omega}}^{\kappa}(\epsilon)\right)^2 = 4\cdot SEF(\epsilon) \left(\sigma_{\epsilon}^{\kappa}\right)^2$.
We infer, therefore, that 
the uncertainty due to fluctuations of the coupling 
strength sets the floor level for our measurements. At the same time, we are pointing out, that this type of noise is not dictated by the 
presence of the exceptional points. This is not the case with the current optical EP-based lasing platforms whose noise-floor is inherently 
associated with the formation of EP itself.

Next we estimate the uncertainty $\sigma_{\epsilon}^{\kappa}$ pertaining to the above analysis. It can be decomposed to the 
thermal noise $\sigma_{\epsilon,th}^{\kappa}$ associated with the Brownian motion of the test-mass and the remaining noise-
sources. We will refer to these sources as added noise sources and they will be characterized by $\sigma_{\epsilon,add}^{\kappa}$. 
Typically, these noise sources prohibit our system from reaching the floor noise level which is dictated by Brownian thermal 
noise. The uncertainty associated with the latter one is: 
\begin{equation}
\left(\sigma_{\epsilon,th}^{\kappa}\right)^2 = \left(\frac{\partial{\epsilon}}{\partial{a}}\right)^2\cdot \alpha^2_{th}\Delta_f,
\label{S36}
\end{equation}
where $a$ is the applied acceleration, and $\alpha_{th}=\sqrt{\frac{4k_bT\omega_n}{mQ}}$ is the thermal noise equivalent 
acceleration (see main text), where $\omega_n = 2\pi\cdot f_n$ and $f_n$ is the natural frequency of the mechanical degree 
of freedom (spring-mass) of our platform, $m$ is the mass of the spring-mass and $Q$- is the quality factor of the mechanical 
spring-mass resonator. The sensitivity parameter $\partial\epsilon/\partial a = Const$ that appears in Eq. (\ref{S36}) describes 
the sensitivity of the coupling strength to the applied acceleration and is determined by the design of the electronic circuit and the spring-mass.

\section{Cumulative characterization of uncertainty in the presence of noise sources}
We are now ready to describe in a compact manner all the above noise sources. The cumulative uncertainty is given as a sum 
of the partial uncertainties i.e. $\sigma_{\Delta \widetilde{\omega}}^2=\left(\sigma_{\Delta\widetilde{\omega}}^{C}\right)^2+\left(
\sigma_{\Delta\widetilde{\omega}}^{\kappa}\right)^2$. Using Eqs. (\ref{S13},\ref{S27},\ref{S30},\ref{S36}) we have
\begin{equation}
\begin{array}{cccc}
\sigma_{\Delta\widetilde{\omega}}^2(\epsilon) \propto 4\Delta_f \left( \frac{k_BT\Gamma^2(\epsilon)}
{|\mathbf{a}_{L1}^{in}|^2}\right )
\left( 2R\frac{\gamma_0}{\gamma_e}\cdot \left(1-\frac{\gamma_0\gamma_e}{(\gamma_0+\epsilon)^2} \right) 
+  Z\cdot \left(1+\frac{\gamma_0^2}{(\gamma_0+
\epsilon)^2} \right)\right) + \\
+ 4\Delta_f\cdot \frac{(\gamma_0+\epsilon)^2}{2\gamma_0\epsilon+\epsilon^2-\gamma_e^2}
\left(\frac{4k_bT\omega_n}{mQ}\left(\frac{\partial{\epsilon}}{\partial{a}}\right)^2+
\left(\sigma_{\epsilon,add}^{\kappa}\right)^2\right)
\end{array}
\label{S38}
\end{equation}

At this point, it is possible to estimate the noise equivalent acceleration $a_{NEA}^2(\epsilon) = {\frac{1}{\Delta_f\cdot\chi^2
(\epsilon)}}\sigma^2_{\Delta\widetilde{\omega}}(\epsilon)$, where the sensitivity of the frequency splitting between the 
transmission peaks with respect to an applied acceleration $a$ is $\chi^2(\epsilon)= 4SEF(\epsilon)\cdot
(\partial{\epsilon}/\partial{a})^2$. We have 
\begin{equation}
\begin{array}{cccc}
\alpha_{NEA}^2(\epsilon) \propto  \left( \frac{k_BT\Gamma^2(\epsilon)}
{|\mathbf{a}_{L1}^{in}|^2}\right )\cdot\left(\frac{\partial{\epsilon}}{\partial{a}}\right)^{-2}\cdot 
\left( 2R\frac{\gamma_0}{\gamma_e}\cdot \left(\frac{(\gamma_0+\epsilon)^2-\gamma_0\gamma_e}{\left(2\gamma_0\epsilon+\epsilon^2-\gamma_e^2\right)^{-1}}  \right) 
+  Z\cdot \left(\frac{(\gamma_0+\epsilon)^2+\gamma_0^2}{\left(2\gamma_0\epsilon+\epsilon^2-\gamma_e^2\right)^{-1}}  \right)\right) + \\
+ \frac{4k_bT\omega_n}{mQ}+
\left(\sigma_{\epsilon,add}^{\kappa}\right)^2\left(\frac{\partial{\epsilon}}{\partial{a}}\right)^{-2}
\end{array}
\label{S39}
\end{equation}
where the first term describes $\alpha_{\cal PT}^2$, the second is $\alpha_{TL}^2$, the third is $\alpha_{th}^2$ and the last term is associated with 
$\alpha_{add}^2$ (see Eq. (5) of the main text).

\begin{figure}
\centering
\includegraphics[width=0.9\columnwidth]{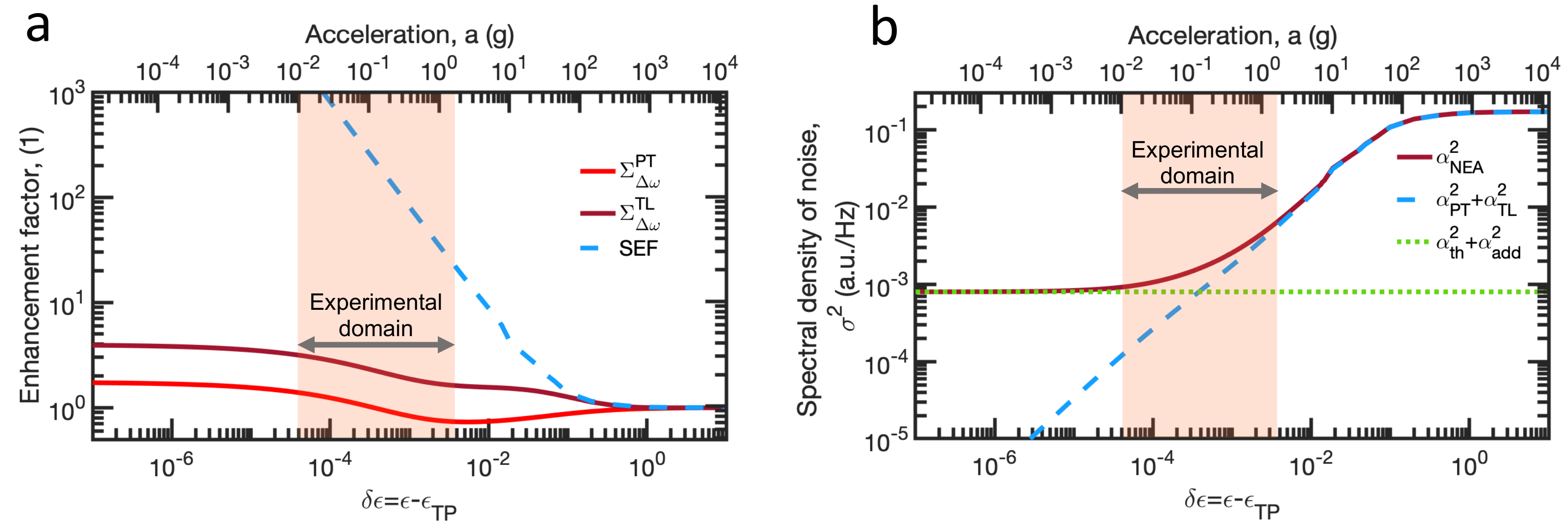}
\caption{\label{figs3} \linespread{0.5}\selectfont{} 
{\bf Noise at the vicinity of $\epsilon=\epsilon_{\rm TPD}$. } {\bf a.}  Enhancement of the uncertainty of the measured transmission peaks splitting at the 
vicinity of $\epsilon=\epsilon_{\rm TPD}$ due to internal noise sources $\left( \Sigma_{\Delta\widetilde{\omega}}^{\cal PT} \right)^2$ (light red  solid line) and external 
noise sources $\left(\Sigma_{\Delta\widetilde{\omega}}^{TL}\right)^2$ (dark red solid line). The sensitivity enhancement factor is shown by blue dashed solid line.  
{\bf b.} Suppression of the uncertainty in the measured acceleration due to the internal and external noise sources (dashed blue line). The red solid line
is the spectral density of the noise variance in the measured acceleration (noise equivalent acceleration $\alpha_{NEA}^2$) as a function of the proximity 
to the $\epsilon_{\rm TPD}$. The spectral density of the variance in the measured acceleration due to the internal and the external noise sources 
($\alpha_{\cal PT}^2+\alpha_{TL}^2$) is shown by a blue dashed line, while the sum of the thermal noise $\alpha_{th}^2$ and the added noise 
$\alpha_{add}^2$ is shown by a green dotted line. In the calculations we have assumed that $f_0=2.68$MHz, $\gamma=0.16$ MHz , $\gamma=0.02$ 
MHz, $\frac{\partial\epsilon}{\partial a }=0.0082$ MHz/g, $R=26\cdot Z$ and that $ \left(\alpha_{th}^2+\alpha_{add}^2\right)\cdot(\frac{\partial\epsilon}{\partial a})^2=0.0002\cdot\frac{4k_BTZ}{|\mathbf{a}_{{\rm TL}_1}^{in}|^2}$}
\end{figure}

From Eq. (\ref{S39}) we can easily deduce that when $\epsilon=\epsilon_{\rm TPD} = -\gamma_0+\sqrt{\gamma_0^2+\gamma_e^2}$  the 
noise contributions $\alpha_{\cal PT}^2+\alpha_{TL}^2$ to $\alpha_{NEA}$
go to zero (see blue dashed line on Fig. \ref{figs3}b). Such behavior indicates that the proposed sensing platform is capable of mitigating these noise effects at the vicinity 
of $\epsilon=\epsilon_{\rm TPD}$. Consequently, at the vicinity of $\epsilon=\epsilon_{\rm TPD}$, the noise equivalent acceleration $a_{NEA}$ is suppressed (see dark red 
line on Fig. \ref{figs3}b) to the value determined by the joint contributions of the thermal noise equivalent acceleration $\alpha_{th}$ and the added noise $\alpha_{add}^2$ 
(see green dotted line on Fig. \ref{figs3}b). Specifically, we will have:
\begin{equation}
\alpha_{NEA}^2(\epsilon\rightarrow \epsilon_{\rm TPD}) = \alpha_{th}^2 +\alpha_{add}^2= \frac{4k_bT\omega_n}{mQ}+ \left(\sigma_{\epsilon,add}^{\kappa} \right)^2  \left(\frac{\partial{\epsilon}}{\partial{a}}\right)^{-2}.
\label{S40}
\end{equation}

\newpage

\end{document}